\documentclass[10pt,conference]{IEEEtran}

\usepackage{url}
\usepackage{tabularx}
\usepackage{adjustbox}
\usepackage{tcolorbox}
\usepackage{caption}
\usepackage{subcaption}
\usepackage{pifont}
\usepackage{enumitem}
\usepackage{makecell}
\usepackage{multirow}
\usepackage{threeparttable}
\usepackage{xcolor}
\usepackage{listings}
\usepackage{afterpage}
\usepackage{placeins} 
\usepackage{float}
\usepackage{graphicx}
\usepackage{booktabs}
\usepackage{amsmath}   % Load first
\usepackage{amssymb}   % Optional, if using extra symbols
\usepackage[colorlinks=true,linkcolor=blue,citecolor=blue,urlcolor=blue]{hyperref}
\usepackage[nameinlink]{cleveref}  % Load after amsmath
\tcbuselibrary{listings, skins, breakable}

\definecolor{highcolor}{RGB}{204,255,204} % Light green for highest values
\definecolor{lowcolor}{RGB}{255,204,204} % Light red for lowest values
\definecolor{codegreen}{rgb}{0,0.6,0}
\definecolor{codegray}{rgb}{0.5,0.5,0.5}
\definecolor{codepurple}{rgb}{0.58,0,0.82}
\definecolor{backcolour}{rgb}{0.95,0.95,0.92}

\lstdefinestyle{prompt}{
    commentstyle=\color{codegreen},
    keywordstyle=\color{magenta},
    stringstyle=\color{codepurple},
    basicstyle=\ttfamily\footnotesize,
    breakatwhitespace=false,         
    breaklines=true,                 
    captionpos=b,
    keepspaces=false,               
    showspaces=false,                
    showstringspaces=false,
    showtabs=false,                  
    tabsize=2,
    literate={\Ü}{{\"U}}1
             {\ä}{{\"a}}1
             {\ö}{{\"o}}1
             {\ü}{{\"u}}1
             {\≤}{{$\leq$}}1
}

\lstdefinestyle{code}{
    backgroundcolor=\color{backcolour},   
    commentstyle=\color{codegreen},
    keywordstyle=\color{magenta},
    numberstyle=\tiny\color{codegray},
    stringstyle=\color{codepurple},
    basicstyle=\ttfamily\footnotesize,
    breakatwhitespace=false,         
    breaklines=true,                 
    captionpos=b,                    
    keepspaces=true,                 
    numbers=left,                    
    numbersep=5pt,                  
    showspaces=false,                
    showstringspaces=false,
    showtabs=false,                  
    tabsize=2
}

\usepackage{pgfplots}
\usepackage{pgfplotstable}
\pgfplotsset{compat=1.18}

\usepackage{colortbl}
\usepackage{ifthen}

\newboolean{showcomments}
\setboolean{showcomments}{false}

\ifthenelse{\boolean{showcomments}}
  {\newcommand{\nb}[3]{{\color{#2}#1: #3}}}
  {\newcommand{\nb}[3]{}}

{\newcommand{\mynote}[2]{
\fbox{\bfseries\sffamily\scriptsize#1}
{\small$\blacktriangleright$\textsf{\emph{#2}}$\blacktriangleleft$}}}
{ \newcommand{\mynote}[2]{}}

\def\BibTeX{{\rm B\kern-.05em{\sc i\kern-.025em b}\kern-.08em
    T\kern-.1667em\lower.7ex\hbox{E}\kern-.125emX}}

\newcommand{\find}[1]{
\begin{tcolorbox}[leftrule=0.4mm,rightrule=0mm,toprule=0mm,bottomrule=0mm,left=0.0pt,right=0.0pt,top=0pt,bottom=0pt]
\em #1
\end{tcolorbox}
}

\begin{document}
\title{Programming Language Confusion: When Code LLMs Can't Keep their Languages Straight}

\author{
\IEEEauthorblockN{
Micheline Bénédicte Moumoula, %\IEEEauthorrefmark{2},
Serge Lionel Nikiema\\
Abdoul Kader Kaboré,
Jacques Klein,
Tegawendé F. Bissyandé
}
\IEEEauthorblockA{
University of Luxembourg, Luxembourg\\
\{micheline.moumoula, lionel.nikiema, abdoulkader.kabore, jacques.klein, tegawende.bissyande\}@uni.lu
}}

\maketitle
\begin{abstract}
Large Language Models (LLMs) have achieved state-of-the-art performance across software engineering tasks, from code generation to translation. However, we identify and systematically evaluate a critical failure mode: Programming Language Confusion (PLC)---the generation of code in unintended languages despite explicit instructions. Through evaluation of 10 popular LLMs across six multilingual datasets (LiveCodeBench, BabelCode variants, HumanEval-XL, and McEval), we demonstrate that PLC is pervasive, with some specialized models exhibiting the highest confusion rates.

Our analysis reveals that PLC is not random noise but reflects systematic patterns: models consistently generate syntactically valid code even when it deviates from language specifications. This behavior produces distinct language migration patterns, most notably a strong default to Python and systematic shifts between syntactically similar language pairs (e.g., C\#/Java). These migrations reflect statistical preferences learned from training data rather than goal-directed reasoning. We demonstrate that explicit language keywords provide the most effective mitigation, while natural language instructions have limited influence on model behavior. Furthermore, model quantization---though essential for practical deployment—significantly amplifies PLC and degrades syntactic stability in complex tasks.

Our findings underscore that language fidelity should be treated as a core evaluation dimension for code LLMs.
We advocate for standardized benchmarks and prompt formats with explicit language constraints to enable more reliable assessment and foster the development of robust, multilingual code generation systems.

\end{abstract}

\begin{IEEEkeywords}
Code generation, Code translation, Code confusion, Large Language Model
\end{IEEEkeywords}

\section{Introduction}
\label{sec:introduction}
Beyond their original text generation function, Large Language Models (LLMs) now operate as sophisticated computational tools, achieving benchmark-leading performance across multiple technical domains, including language understanding, formal reasoning, code synthesis, and other diversely complex software engineering tasks~\cite{kaddour2023challenges}. 
In particular, code-specialized language models (a.k.a code LMs) have emerged as significant implementations of foundation models~\cite{anil2023palm, awais2025foundation}, with substantial empirical evidence documenting their integration throughout the software development lifecycle~\cite{barke2023grounded, bird2022taking}. These models have been deployed across multiple development environments~\cite{sergeyuk2025using} and function within constrained IDE contexts~\cite{ziegler2022productivity}, as automated programming aids~\cite{xu2022systematic}, and as components in complex agent systems~\cite{wang2023survey}. Studies have demonstrated their ability to automate repetitive coding tasks~\cite{10.1145/3597503.3639187, 10.1145/3597503.3639219, chen2024survey, liu2023your}, potentially reducing implementation time~\cite{peng2023impact}, augmenting programmer efficacy~\cite{dakhel2023github}, and expanding access to complex programming techniques for developers with varied skill levels~\cite{vaithilingam2022expectation}.

As the adoption of AI-generated code in production environments accelerates, foundation models are presenting subtle reliability challenges that demand rigorous investigation~\cite{bommasani2021opportunities}. 
%These collective efforts underscore the critical challenge facing the field: leveraging the productivity and innovation gains of automated code generation without compromising on the quality, reliability, and security of software systems.
Indeed, the integration of LLM-generated code into critical systems demands unprecedented levels of correctness, security, and robustness.
Rigorous investigations into the phenomenon of ``hallucination''--where models confidently produce non-functional or even software-unrelated outputs--have revealed concerning patterns that demand immediate attention.
The research community has therefore developed diverse benchmarks to evaluate the functional accuracy and safety of code generation systems under realistic constraints~\cite{chen2021codex, Austin2021ProgramSW}.
In parallel, researchers have constructed comprehensive taxonomies~\cite{liu2024exploring, pan2024lost} that categorize the wide spectrum of programming errors generated by Code LMs, ranging from basic syntax and logic issues to sophisticated security vulnerabilities and edge-case failures.
These collective efforts underscore the critical challenge facing the field: how to leverage the productivity and innovation gains of LLM-driven automated code generation without compromising the quality, reliability, and security of software systems.

{\bf This paper.} We investigate a surprising limitation of LLMs in the domain of software engineering: \textit{their inability to consistently generate code in a user's desired programming language}, as exemplified in Listing~\ref{lst:example}. This limitation poses fundamental concerns beyond anecdotal failures since inadvertent confusion between two languages may introduce subtle, difficult-to-detect errors that compromise both reliability and security.

% Style for all code blocks
\lstdefinestyle{codeblock}{
    backgroundcolor=\color{gray!15},  % colored background for the code
    basicstyle=\ttfamily\small,
    keywordstyle=\color{blue}\bfseries,
    commentstyle=\color{green!50!black}\itshape,
    stringstyle=\color{red!70!black},
    numberstyle=\tiny\color{gray},
    numbers=none,
    frame=none,
    breaklines=true,
    tabsize=4,
    postbreak=\mbox{\textcolor{red}{$\hookrightarrow$}\space},
}

% Python-specific style for output
\lstdefinestyle{pythonblock}{
    language=Python,
    backgroundcolor=\color{gray!5},
    basicstyle=\ttfamily\small,
    keywordstyle=\color{blue}\bfseries,
    commentstyle=\color{green!50!black}\itshape,
    stringstyle=\color{red!70!black},
    numberstyle=\tiny\color{gray},
    numbers=none,
    frame=none,
    breaklines=true,
    tabsize=4,
    postbreak=\mbox{\textcolor{red}{$\hookrightarrow$}\space},
}

% The figure

\begin{figure}[!h]
\centering
\begin{tcolorbox}[
    colframe=black!70,
    colback=white,
    sharp corners,
    enhanced,
    %breakable,
    left=2mm,
    right=2mm,
    top=1mm,
    bottom=1mm,
    boxsep=1mm
]

\textbf{Prompt:}
\begin{lstlisting}[style=codeblock]
class Solution {
    /**
     * Write a java function to find the last two digits in factorial of a given number.
     */
    public Integer lastTwoDigits(Integer n) {
\end{lstlisting}

\tcblower

\textbf{Output of CodeLLama-13B-Instruct-Q:}
\begin{lstlisting}[style=pythonblock, label={lst:example}, caption={Programming language confusion example}, captionpos=b]
def get_last_two_digits_of_factorial(n):
    fact = 1
    for i in range(2, n+1):
        fact *= i
    return str(fact)[-2:]
# Test cases
assert get_last_two_digits_of_factorial(1) == "1"
assert get_last_two_digits_of_factorial(10) == "4"
assert get_last_two_digits_of_factorial(100) == "7"
\end{lstlisting}

\end{tcolorbox}
\vspace{-0.5cm}
\end{figure}

Our study investigates to what extent programming language confusion occurs across 10 LLMs on six datasets related to code generation and translation tasks, involving about 75k program samples written in 16 programming languages. A key finding of our study is that \textit{\textbf{programming language confusion in LLMs often reflects systematic output preferences rather than random mistakes}}: models frequently migrate to languages in which they can produce more syntactically valid code. These migration patterns (for example, frequent moves toward Python and between syntactically similar language pairs) are consistent with statistical preferences in the learned model distribution, and should not be interpreted as evidence of deliberate or goal-directed behavior by the models.

Our contributions are as follows:

\begin{itemize}[leftmargin=*]
  \item We conduct the first comprehensive, large-scale empirical investigation of programming language confusion in LLMs, systematically analyzing code generation and translation tasks across multiple programming languages and natural languages.

  \item We quantify the impact of model quantization on PLC, demonstrating that this necessary deployment step significantly amplifies confusion and compromises the syntactic stability of model outputs.
  
  \item We provide novel insights into the factors contributing to programming language confusion, revealing critical limitations in current LLM design that affect multilingual code generation reliability.
  
\end{itemize}

\section{Background \& Related Work}
\label{sec:background}
This section contextualizes our investigation of programming language confusion within the broader landscape of code language models and reliability challenges in AI-assisted software engineering.

\subsection{Code Language Models}
Code language models have evolved rapidly since the introduction of early approaches such as code2vec~\cite{alon2019code2vec} and CodeBERT~\cite{feng2020codebert}. Modern code LMs like Codex~\cite{chen2021codex}, AlphaCode~\cite{li2022competition}, PaLM-Coder~\cite{anil2023palm}, and Code Llama~\cite{roziere2023code} leverage large-scale transformer architectures trained on vast code repositories to generate functionally correct code from natural language specifications. These models typically employ similar pretraining objectives as general-purpose LLMs but with specialized training data composed of code from various programming languages~\cite{xu2022systematic}.
The effectiveness of code language models depends significantly on their representational capacity for programming languages. Unlike natural languages, programming languages exhibit strict syntactic constraints and semantic rules that must be followed precisely for code to execute correctly~\cite{yin2017syntactic}. This presents unique challenges for language models, particularly in the context of multiple programming languages where syntax and paradigms can vary substantially~\cite{ahmad2021unified}. Research by Xu et al.~\cite{xu2022systematic} demonstrates that these models can learn multiple programming languages simultaneously, but the mechanisms governing cross-language interaction and potential interference remain underexplored.

\subsection{Reliability Challenges in LLM-Generated Code}
As LLM-generated code transitions from research settings to production environments, reliability has emerged as a critical concern. Recent studies have documented various shortcomings in code generation capabilities, including syntactic errors~\cite{liu2024exploring}, semantic inconsistencies~\cite{pan2024lost}, security vulnerabilities~\cite{pearce2022asleep}, and hallucinated API references~\cite{chen2021evaluating}.
Taxonomies of errors in LLM-generated code have been proposed by Liu et al.\cite{liu2024exploring} and Pan et al.\cite{pan2024lost}, categorizing issues from basic syntax errors to complex logical flaws. However, these taxonomies do not specifically address programming language confusion as a distinct category of error. The economic impact of such errors has been analyzed by Sobania et al.~\cite{sobania2022choose}, who found that fixing AI-generated code defects can sometimes exceed the effort saved by the initial code generation.
Various benchmarks have been developed to evaluate code LMs, including HumanEval~\cite{chen2021codex}, MBPP~\cite{austin2021program}, and CodeXGLUE~\cite{lu2021codexglue}, which test functional correctness across different programming tasks. More recently, multi-language benchmarks like MultiPL-E~\cite{cassano2022multipl} have enabled cross-linguistic comparisons, but they typically evaluate each language in isolation rather than examining cross-language interference explicitly.

\subsection{Cross-language Phenomena in Language Models}
While extensive research exists on multilingual capabilities in natural language models~\cite{conneau2020unsupervised}, similar investigations in programming languages remain limited. The phenomenon of code-switching and language mixing in natural language has been studied by Winata et al.~\cite{winata2021language}, who found that language models can struggle with maintaining linguistic boundaries during generation. Analogous behaviors in code generation contexts would have significant implications for software reliability.
Cross-lingual transfer learning has been explored by Ahmad et al.\cite{ahmad2021unified} and Wang et al.\cite{wang2021codet5}, demonstrating that models can leverage syntactic similarities between programming languages. However, these studies focus primarily on leveraging cross-language knowledge for improved performance rather than investigating potential negative interference between languages.

\subsection{Human Programming Language Cognition}
Research in human cognition provides relevant context for understanding language confusion in LLMs. Studies by Kahneman~\cite{kahneman2011thinking} on dual-system thinking and cognitive load theory applied to programming by Siegmund et al.\cite{siegmund2014understanding} suggest that context switching between programming languages imposes significant cognitive demands. Human programmers develop explicit strategies to manage these demands and avoid confusion between languages with similar syntax\cite{lee2013knowledge}.
Peitek et al.~\cite{peitek2021program} have used fMRI studies to demonstrate that different programming languages activate distinct neural patterns in human programmers, suggesting specialized cognitive processes for language-specific comprehension. Whether LLMs exhibit analogous specialization or more fluid boundaries between language representations remains an open question.

\subsection{Position of Our Work}
Our research differs from previous work in several key respects. Unlike studies that evaluate models on individual programming languages in isolation~\cite{chen2021codex, austin2021program}, we specifically investigate cross-language confusion and interference. While taxonomies of LLM coding errors~\cite{liu2024exploring, pan2024lost} provide useful frameworks, they do not specifically address language confusion as a distinct error category—a gap our work aims to fill.
Most closely related to our research is work by Cassano et al.\cite{cassano2022multipl} on multi-language evaluation, but their focus remains on comparative performance rather than systematic analysis of cross-language confusion patterns. Similarly, while Ahmad et al.\cite{ahmad2021unified} explore cross-lingual transfer between programming languages, they do not investigate negative transfer or confusion.
Our systematic evaluation across multiple languages and models, coupled with our analysis of contributing factors and proposed mitigation strategies, represents a novel contribution to understanding the reliability challenges of LLM-based code generation in multilingual contexts.
\section{Experimental Setup}
\label{sec:setup}
\begin{figure*}[!ht]
        \centering
        \includegraphics[width=0.8\linewidth]{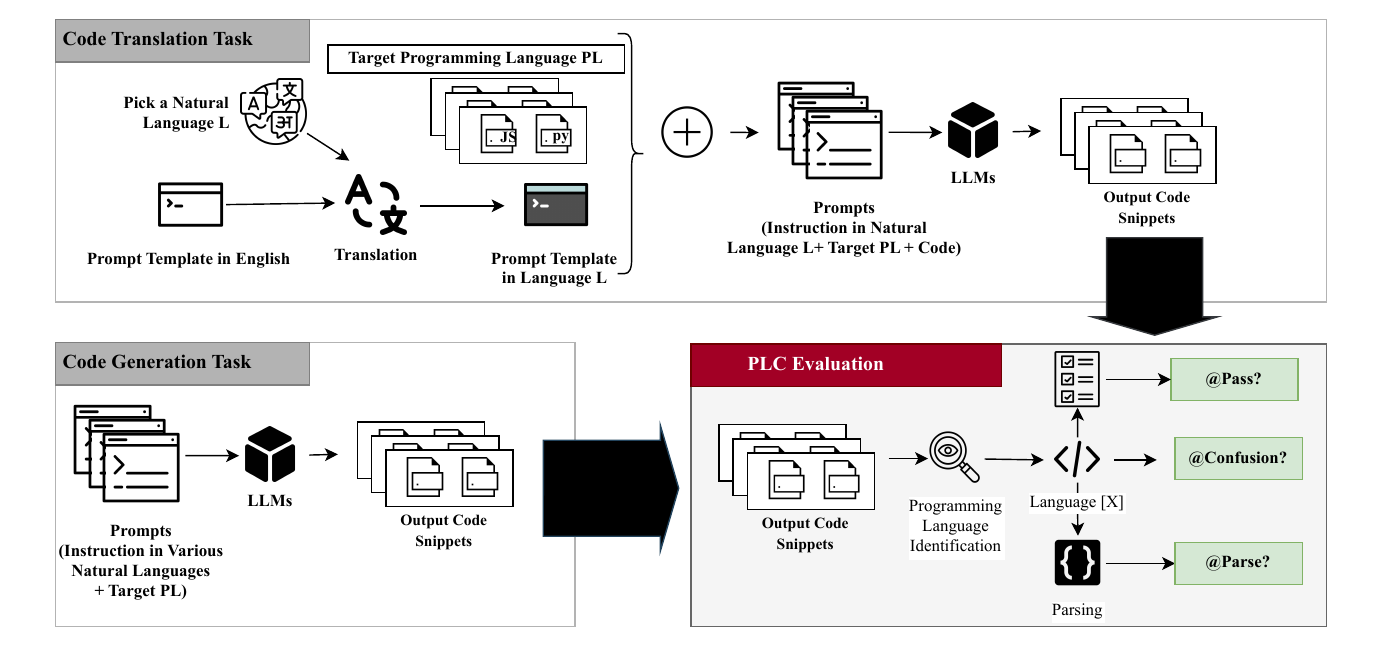} 
    \vspace{-0.2cm}    
    \caption{Experimental workflow for programming language evaluation}
    \label{fig:workflow}
\end{figure*}

This section details our experimental methodology for investigating programming language confusion in LLMs, including our research questions, model selection criteria, datasets, evaluation metrics, and procedural workflow.

\subsection{Research Questions} \label{rqs}
Our investigation focuses on three main research questions.
\begin{itemize}[leftmargin=*]
    \item \textbf{RQ1}:  {\em To what extent does programming language confusion manifest in LLM-based code generation?} We quantify the prevalence of language confusion in code generation tasks and analyze its correlation with code correctness across different programming languages and model architectures.

    \item \textbf{RQ2}: {\em What factors influence programming language confusion during LLM-based code translation?} We investigate whether LLMs maintain target language adherence when explicitly tasked with translating code between programming languages, and examine how translation direction and language pairs affect confusion patterns.
    
    \item \textbf{RQ3}: {\em How does model quantization affect programming language confusion in LLM-based code tasks?} We evaluate whether quantized models exhibit different levels of language confusion or code correctness compared to their full-precision counterparts across the generation and translation settings.

\end{itemize}

\subsection{Models}
\label{lab:models}
We evaluate 10 LLMs spanning both general-purpose and code-specialized architectures to enable comparative analysis of language confusion tendencies. Our selection includes:

\begin{itemize}[leftmargin=*]
    \item \textbf{General-purpose LLMs}: DeepSeek-V2-Lite-Instruct-16B~\cite{zhu2024deepseek},  GPT-3.5-Turbo~\cite{achiam2023gpt}, GPT-4.1-Mini~\cite{openai2024gpt4technicalreport}, LLama3.1-8B-instruct~\cite{grattafiori2024llama}, Mistral-7B-Instruct~\cite{jiang2310mistral}, Qwen2.5-Instruct-14B~\cite{qwen2.5}
    
    \item \textbf{Code-specialized LLMs}: CodeLLama-13B-Instruct~\cite{roziere2023code}, DeepSeek-Coder-V2-Lite-Instruct-16B~\cite{zhu2024deepseek}, Qwen2.5-Coder-Instruct-14B~\cite{qwen2.5}, Starcoder2-15B-Instruct~\cite{wei2024selfcodealign}.
\end{itemize}
Model selection criteria include architectural diversity, parameter scale variation (7B to 16B parameters), demonstrated performance on coding benchmarks~\cite{xu2022systematic, liu2023your}.
For all experiments we used greedy decoding (temperature $T=0$) to mitigate randomness while preserving output coherence, following established practices in the code-generation literature~\cite{chen2021evaluating}. To elicit full, self-contained solutions, we appended a small instruction prefix to the original benchmark prompt (see Appendix).

\subsection{Datasets}
\label{lab:datasets}

To comprehensively assess language confusion across diverse contexts, we selected established multilingual datasets for both code generation and translation tasks:

\begin{itemize}[leftmargin=*]
    \item \textbf{Code Generation Datasets}:
    \begin{itemize}[leftmargin=*]
        \item \textbf{LiveCodeBench}~\cite{jain2024livecodebench}: A holistic, contamination-free benchmark using time-tagged competitive programming problems to evaluate code LLMs across generation, debugging, and execution tasks.

        \item \textbf{BabelCode-HumanEval}~\cite{orlanski2023measuring}: A cross-lingual extension of the HumanEval benchmark~\cite{chen2021codex} spanning 16 programming languages, with consistent function signatures, documentation standards, and test cases across languages.
        
        \item \textbf{BabelCode-MBPP}: A multilingual adaptation of the MBPP dataset~\cite{Austin2021ProgramSW} containing approximately 1,000 basic programming problems with natural language descriptions and test cases.
        
        \item \textbf{BabelCode-TP3}: Derived from Python Programming Puzzles~\cite{schuster2021programming}, this dataset features verification functions of varying complexity, from string manipulation to advanced algorithms.
        
        \item \textbf{HumanEval-XL}~\cite{peng2024humaneval}: An extension of HumanEval with problem statements translated to multiple natural languages, enabling evaluation of cross-linguistic influence on code generation.
    \end{itemize}
    
    \item \textbf{Code Translation Dataset}:
    \begin{itemize}[leftmargin=*]
        \item \textbf{McEval}~\cite{chai2024mcevalmassivelymultilingualcode}: A comprehensive benchmark for evaluating code generation and understanding, containing diverse programming tasks and problem descriptions, supporting structured assessment of translation capabilities between programming language pairs.

    \end{itemize}
\end{itemize}

\begin{table}[!h]
    \centering
\caption{Summary information on experimental datasets}
\label{tab:dataset}
\renewcommand{\arraystretch}{1.4}
\centering
          \resizebox{0.9\linewidth}{!}{
                \begin{tabular}{c|c|c|c|c}
                \hline
                   \bf  Task  &\bf  Instruction language & \bf Dataset name & \bf Languages & \bf \# of samples \\ \hline
                      \multirow{4}{*}{\vspace{-0.1cm}Code generation} & \multirow{3}{*} {\vspace{-0.1cm}English}
                      &  LiveCodeBench&  1 PL & 880 \\
                      &&  BC-TP3&  13 PLs & 5920 \\
                      &&  BC-MBPP&  13 PLs & 13673 \\
                      &&  BC-HumanEval& 13 PLs  & 2576\\  \cline{2-5} 
                      &23 NLs&  HumanEval-XL& 12 PLs  & 22080 \\ \hline
             Code translation& 23 NLs & McEval & 14 PLs & 33488 \\ \hline
    \end{tabular}}
    \vspace{-0.4cm}
\end{table}

Dataset selection was guided by three criteria: (1) coverage of diverse programming paradigms and language families, (2) availability of parallel implementations across languages, and (3) established usage in prior code generation research~\cite{xu2022systematic, chen2021evaluating, chen2021codex}.

\subsection{Evaluation Metrics}
\label{lab:metrics}
We define three complementary metrics to evaluate the extent of language confusion:

\begin{itemize}[leftmargin=*]
    \item \textbf{Language Confusion Pass Rate (LCPR)}: Measures the proportion of the generated code samples that correctly match the target programming language. Formally defined as:
    \begin{equation}
        \text{LCPR} = \frac{\text{Number of samples in correct language}}{\text{Total number of samples}}
    \end{equation}
    This metric, which builds on precision, quantifies a model's ability to maintain language boundaries during generation, i.e. whether it can accurately generate code in the expected programming language. The determination of the language, given a code, is done following the methodology described in Section~\ref{lab:methodology}.

    \item \textbf{Code Parsing Pass Rate (CPPR)}: Assesses the syntactic validity of generated code by measuring the proportion of samples that parse without errors in the detected language:
    \begin{equation}
        \text{CPPR} = \frac{\text{Number of samples that parse successfully}}{\text{Total number of samples}}
    \end{equation}

    \item \textbf{Dominant Migration Rate (DMR)}: Measures the proportion of language-confused samples that migrate to a specific target language (e.g., Python):
     \begin{equation}
         \text{DMR}(X) = \frac{\text{Number of confused samples in language } X}{\text{ Total number confused samples}}
     \end{equation}
\end{itemize}
These metrics are based on well-established principles from compiler theory and multilingual NLP evaluation~\cite{conneau2020unsupervised}, tailored specifically to analyze programming language confusion.

\subsection{Methodology}
\label{lab:methodology}
Our experimental workflow, illustrated in Figure~\ref{fig:workflow}, consists of two primary evaluation streams for code generation and translation tasks:
\begin{itemize}[leftmargin=*]
    \item \textbf{Code Generation Task}: We employ a three-stage process:
    \begin{enumerate}
        \item \textit{Language Model Inference}: Each model generates code responses to prompts from the five code generation benchmarks.
        
        \item \textit{Programming Language Detection}: We implement an ensemble approach using three specialized tools to reliably identify the programming language of generated code: 
        %\noindent
        \begin{itemize}[leftmargin=*]
            \item \texttt{Highlight.js\footnote{https://highlightjs.org/}}: A widely-used syntax detection library supporting 192 programming languages.
            \item \texttt{Guesslang\footnote{https://guesslang.readthedocs.io/}}: A deep learning-based language identification system with $>90\%$ reported accuracy across 50+ languages.
            \item \texttt{Philomath-1209\footnote{https://huggingface.co/philomath-1209/programming-language-identification}}: A fine-tuned CodeBERTa model specifically designed for programming language identification.
            \item \texttt{PLangRec~\cite{RODRIGUEZPRIETO2025107640}}: A deep-learning model for programming language recognition that leverages a large balanced dataset and ensemble learning to classify code at line, snippet, and file level with state-of-the-art accuracy.

        \end{itemize}
        
        \item \textit{Language Determination Algorithm}: We employ a weighted consensus approach prioritizing Philomath-1209 due to its demonstrated accuracy. When tools disagree, we apply a decision tree that considers language-specific characteristics and the confidence scores of each detector. We validated the pipeline on a subset covering the 16 programming languages used in our study and found the ensemble achieves an overall accuracy of 98.2\%, confirming the reliability of our detection approach; per-language results and full confusion matrices are provided in the supplementary file.
        \item \textit{Evaluation}: The extracted code is then parsed using Tree-sitter\footnote{https://tree-sitter.github.io/tree-sitter/}, an incremental parsing library that generates concrete syntax trees. This allows us to identify syntax errors within the code. Finally, we compare the target programming language with the detected programming language to identify the language confusion, where code is mistakenly written in the wrong language.
    \end{enumerate}
    %\vspace{-0.5cm}
    \item \textbf{Code Translation Task}: We extend our methodology to translation contexts:
    \begin{enumerate}
        \item \textit{Instruction Translation}: Prompt instructions were translated into 23 natural languages using GPT-4 and validated through back-translation to ensure semantic fidelity.
        
        \item \textit{Translation Generation}: LLMs translate source code between language pairs while following instructions in various natural languages.
        
        \item \textit{Language Detection and Validation}: We apply the same detection ensemble and Tree-sitter parsing as in the generation task to identify language confusion instances.
    \end{enumerate}
\end{itemize}

This methodology enables systematic identification and quantification of programming language confusion across diverse contexts, providing the empirical foundation for addressing our research questions.
\section{Evaluation}
\label{sec:evaluation}
This section presents our empirical findings organized by research question. For each research question defined in Section~\ref{rqs}, we first describe the corresponding experiments conducted and then, present quantitative results, discuss and analyze key implications.

\subsection{[RQ1] Programming Language Confusion in LLM-Based Code Generation}
\label{rq1}

\noindent
\textbf{Goal.} We investigate how effectively LLMs maintain programming language fidelity during code generation. Specifically, we quantify the extent of PLC and examine factors that influence it, including language families, model specialization, and instruction language.

\noindent
\textbf{Experiments.} We conducted three sets of experiments:
\begin{enumerate}[leftmargin=*,noitemsep,topsep=2pt]
    \item An evaluation on {LiveCodeBench} using English instructions that omit explicit language specification. This determines the models' default language schema (baseline PLC).
    \item An evaluation on {BabelCode} datasets with explicit English instructions, assessing LCPR, CPPR, and DMR across 13 diverse target languages.
    \item An analysis using {HumanEval-XL} with instructions in 23 natural languages to examine how instruction language affects LCPR.
\end{enumerate}
We also report in this section the Functional Pass Rate (FPR) that measures the proportion of generated programs that execute successfully and pass at least one unit test:
\begin{table}[!h]
    \centering
    \caption{PLC during Code Generation with LiveCodeBench}
    \renewcommand{\arraystretch}{1.2}
    \resizebox{1\linewidth}{!}{
        \small
        \begin{tabular}{l|c|c|c|c}
        \hline
        \textbf{LLMs} & \textbf{LCPR} & \textbf{CPPR non confuse \%} & \textbf{CPPR confuse \%} & \textbf{DMR \%} \\ \hline
        \textbf{CodeLLama-13B-Instruct} & 98.86 & 99.77 &  100.00 & 100 (c++) \\
        %CodeLlamaInstruct-34B & 99.30 & \cellcolor{highcolor} 100.00 & 50.00 \\
        \textbf{DeepSeek-Coder-V2-Lite-Instruct-16B} &  100.00 & 99.89 & - & -\\
        \textbf{DeepSeek-V2-Lite-Instruct-16B} &  99.89 & 99.54 & 0.00 & 100 (scala)\\
        \textbf{GPT-3.5-Turbo} &  100.00 & 99.66 & - & -\\
        \textbf{GPT-4.1-Mini} & 99.89 & 100.00 & 0.00 & 100 (scala)\\
        \textbf{LLama3.1-8B-instruct} & 97.22 & \cellcolor{lowcolor} 91.79 & 87.50 & 91.67 (c++) \\
        \textbf{Mistral-7B-Instruct} & \cellcolor{lowcolor} 66.70 & 99.66 & 93.10 & 93.10 (c++)\\
        \textbf{Qwen2.5-Instruct-14B} & 99.89 & 99.89 & \cellcolor{lowcolor} 0.00 & 100 (scala)\\
        \textbf{Qwen2.5-Coder-Instruct-14B} &  100.00 &  100.00 & - & -\\
        \textbf{Starcoder2-15B-Instruct} &  100.00 & 99.89 & -  & -\\
        \hline
        \end{tabular}
    \label{tab:livecodebench} }
\end{table}

% \begin{table}[!h]
%     \centering
%     \caption{PLC during Code Generation with LiveCodeBench}
%     \renewcommand{\arraystretch}{1.1}
%     \resizebox{1\linewidth}{!}{
%         \small
%         \begin{tabular}{l|c|c|c}
%         \hline
%         \textbf{LLMs} & \textbf{LCPR} & \textbf{CPPR non confuse \%} & \textbf{CPPR confuse \%} \\ \hline
%         CodeLLama-13B-Instruct-Ollama & 97.49 & 99.88 & 81.82 \\
%         LLama3.1-8B-instruct-ollama & 97.22 & 91.79 & 87.50 \\
%         Mistral-7B-Instruct &  99.89 & 97.72 & 100.00 \\
%         Starcoder2-15B-Instruct-Ollama & 99.66 & 99.66 & 100.00 \\ 
%         \hline
%         \end{tabular}
%     \label{tab:livecodebench-o} }
% \end{table}
 % Assuming this is the placeholder for your LiveCodeBench table

\noindent
\textbf{Results.}\subsubsection{Python as the Universal Attractor}
\Cref{tab:livecodebench} presents PLC results on LiveCodeBench, a Python-focused benchmark where instructions deliberately omit explicit language specification. The results demonstrate that most LLMs have internalized a strong Python default, with four models achieving a perfect $100.00\%$ LCPR. This confirms Python as the established default coding language when instructions omit explicit language specification. However, this fidelity is not universal, with Mistral-7B-Instruct exhibiting significant PLC (66.70\% LCPR), often defaulting to C++. A critical finding is the instability in code quality independent of language confusion, evidenced by Llama3.1-8B-Instruct showing reduced syntactic correctness ($91.79\%$ $CPPR_{non-confused}$) even when correctly targeting Python, suggesting a fundamental failure in syntactic generation beyond mere language switching.

\begin{table}[!http]
    \centering
    \caption{PLC in \% during Code Generation with BabelCode (BC)}
    \renewcommand{\arraystretch}{1.3}
    \footnotesize
          \resizebox{1\linewidth}{!}{
                    \begin{tabular}{l|c|c|c|c|l}
                    \hline
                        \textbf{LLMs} & \textbf{\makecell{Dataset}} &\textbf{\makecell{LCPR}} & \textbf{\makecell{CPPR non-confuse} }  & \textbf{\makecell{CPPR confuse}} & \textbf{\makecell{DMR \%}} \\ \hline
                         \multirow{4}{*} {\vspace{-0.1cm} CodeLLama-13B-Instruct }
                         & BC-TP3 & 88.18 & 98.64 & 86.47 & 43.98 (java)\\
                         & BC-MBPP & 87.26 & 97.35 & 95.84 & 36.01 (java) \\
                         & BC-HumanEval & 82.71 & 97.85 & 91.14 & 38.00 (py) \\ \cline{2-6}
                         & All &  86.96 & 97.75 & 92.89 & 37.09 (java) \\ \hline

                         \multirow{4}{*} {\vspace{-0.1cm} \makecell{DeepSeek-Coder-V2-\\Lite-Instruct-16B}} %  \parbox{4cm}{ DeepSeek-Coder-V2-\\Lite-Instruct-16B} }
                         & BC-TP3 & 77.39 & 99.17 & 85.77 & 91.89 (py)\\
                         & BC-MBPP & 91.63 & 99.73 & 97.92 & 62.83 (py)\\
                         & BC-HumanEval & 84.89 & 99.83 & 100.00 & 70.68 (py)  \\ \cline{2-6}
                         & All &  87.06 & 99.61 & 92.54 & 77.41 (py)   \\ \hline

                         \multirow{4}{*} {\vspace{-0.1cm} DeepSeek-V2-Lite-Instruct-16B }
                         & BC-TP3 & 85.23 & 97.92 & 91.07 & 42.75 (py)\\
                         & BC-MBPP & 86.63 & 98.77 & 95.34 & 54.60 (py) \\
                         & BC-HumanEval & 84.36 & 98.14 & 95.61 & 46.08 (py)  \\ \cline{2-6}
                         & All &  86.00 & 98.48 & 94.19 & 50.20 (py)   \\ \hline
                         
                         \multirow{4}{*} {\vspace{-0.1cm} GPT-3.5-Turbo }
                         & BC-TP3 & 98.74 & 99.10 & 71.67 & 76.67 (java)\\
                         & BC-MBPP & 91.73 & 99.56 & 97.34 & 54.94 (java) \\
                         & BC-HumanEval & 92.48 & 98.94 & 98.05 & 66.23 (java)  \\ \cline{2-6}
                         & All &  93.69 & 99.36 & 96.05 & 57.67 (java)  \\ \hline

                         \multirow{4}{*} {\vspace{-0.1cm} GPT-4.1-Mini }
                         & BC-TP3 & 97.65 & 99.28 & 78.38 & 31.53 (java)\\
                         & BC-MBPP & 92.48 & 99.87 & 96.70 & 43.59 (py) \\
                         & BC-HumanEval & 94.70 & 99.95 & 97.22 & 80.56 (java)  \\ \cline{2-6}
                         & All &  94.12 & 99.72 & 94.80 & 45.38 (java)  \\ \hline
                         
                         \multirow{4}{*} {\vspace{-0.1cm} LLama3.1-8B-instruct }
                         & BC-TP3 & 72.28 & 95.13 & 91.23 & 40.56 (js)\\
                         & BC-MBPP & 81.51 & 96.42 & 95.62 & 42.35 (py) \\
                         & BC-HumanEval & 67.78 & 94.64 & 95.06 & 36.27 (js)  \\ \cline{2-6}
                         & All & 77.47 & \cellcolor{lowcolor}95.92 & \cellcolor{lowcolor}94.10 & 37.08 (py)   \\ \hline

                         \multirow{4}{*} {\vspace{-0.1cm} Mistral-7B-Instruct }
                         & BC-TP3 & 93.10 & 98.24 & 95.31 & 87.50 (js)\\
                         & BC-MBPP & 92.55 & 99.33 & 96.67 & 39.21 (java) \\
                         & BC-HumanEval & 87.03 & 98.80 & 96.17 & 55.17 (js)  \\ \cline{2-6}
                         & All &  92.07 & 98.98 & 96.26 & 46.34 (js)   \\ \hline

                         \multirow{4}{*} {\vspace{-0.1cm} Qwen-2.5-14B-Instruct }
                         & BC-TP3 & 99.20 & 99.42 & 92.11 & 65.79 (js)\\
                         & BC-MBPP & 96.93 & 99.54 & 98.51 & 69.85 (java) \\
                         & BC-HumanEval & 97.51 & 99.75 & 98.04 & 62.75 (java)  \\ \cline{2-6}
                         & All &  97.60 & 99.53 & 97.88 & 63.92 (java)   \\ \hline

                         \multirow{4}{*} {\vspace{-0.1cm} Qwen-2.5-Coder-14B-Instruct }
                         & BC-TP3 & 95.30 & 99.71 & 99.02 & 50.20 (py)\\
                         & BC-MBPP & 95.30 & 99.71 & 99.02 & 50.20 (py) \\
                         & BC-HumanEval & 97.51 & 99.80 & 98.04 & 43.14 (java)  \\ \cline{2-6}
                         & All &  96.56 & 99.77 & 99.01 & 50.08 (py)  \\ \hline

                         \multirow{4}{*} {\vspace{-0.1cm} StarCoder2-15B-Instruct }
                         & BC-TP3 & 65.24 & 99.12 & 93.10 & 78.92 (py)\\
                         & BC-MBPP & 78.74 & 99.40 & 96.86 & 56.64 (py) \\
                         & BC-HumanEval & 71.58 & 99.22 & 98.21 & 71.91 (py)  \\ \cline{2-6}
                         & All &  \cellcolor{lowcolor}74.33 & 99.32 & 95.68 & 66.59 (py)   \\ \hline
                    \end{tabular}}
         \label{tab:babelcode}
\vspace{-0.4cm}
\end{table}

\noindent
\subsubsection{Quantifying PLC (BabelCode)}
\label{results:babel}
\noindent
From \Cref{tab:babelcode}, we observe significant LCPR variation. Qwen2.5-Instruct-14B demonstrates the strongest language adherence at 97.60\% LCPR, while StarCoder2-15B-Instruct exhibits the lowest fidelity at 74.33\% LCPR. Crucially, some general-purpose models (e.g., GPT-3.5-Turbo) rival or outperform code-specialized models (e.g., CodeLLama-13B-Instruct), supporting the finding that code specialization does not guarantee superior language boundary maintenance.

\noindent
\textit{Systematic Language Migration (CPPR).} A striking pattern emerges: syntactic correctness is preserved despite language confusion. Most models achieve $CPPR_{non-confused}$ above $95\%$. Intriguingly, $CPPR_{confused}$ values are often comparable to the non-confused rates. This counterintuitive behavior suggests a systematic language migration where models prioritize structural validity, deliberately defaulting to languages (likely those over-represented in training) where they are more confident in producing syntactically reliable code, even when this contradicts the language request.

\noindent
\textit{Functional Degradation.} Supplementary file Table 2 reveals that high CPPR is a false indicator of quality. 
While confused outputs maintain high CPPR, their FPR degrades substantially.
For instance, Starcoder2-15B-Instruct's FPR drops from $39-76\%$ when correct to $21-63\%$ when confused. 
This shows that PLC introduces semantic errors beyond syntax—models generate code that is parseable but fails unit tests, revealing that systematic language selection optimizes for surface-level correctness at the expense of functional accuracy. However, the fact that models still produce some functionally correct code in confused samples suggests that cross-language semantic understanding is partially preserved despite language boundary violations.

\noindent
\subsubsection{Language Migration Patterns and Biases}
\begin{figure}[h]
\vspace{-0.4cm}
        \centering
    \begin{minipage}{0.49\linewidth}
        \includegraphics[width=0.9\linewidth]{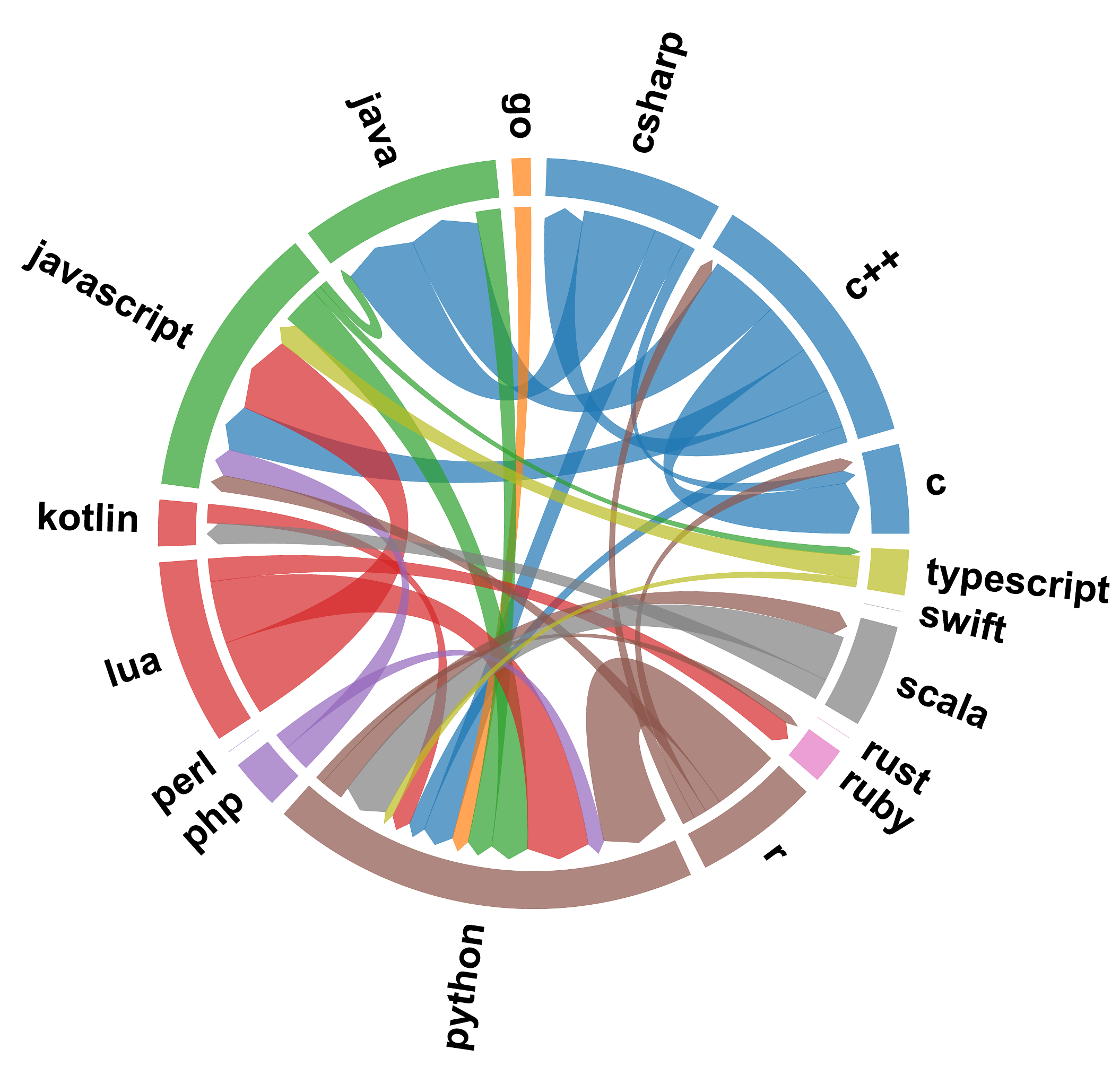}
        \centering
        \textbf{\footnotesize (a)~CodeLLama-13B-Instruct}
    \end{minipage}
    \begin{minipage}{0.49\linewidth}
        \includegraphics[width=0.9\linewidth]{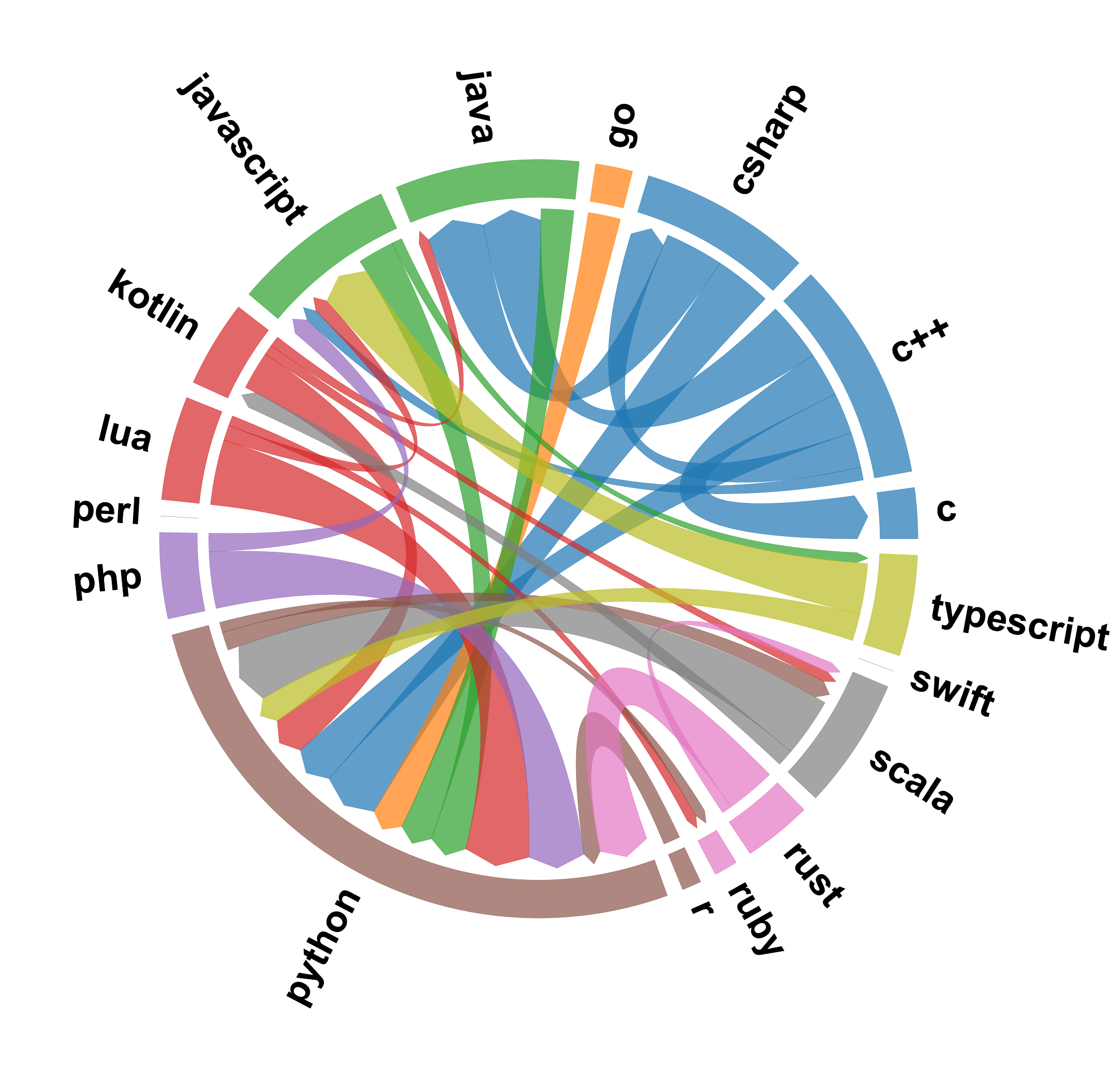}
        \centering
        \textbf{\footnotesize (b)~Starcoder2-15B-Instruct}
    \end{minipage}
    
    \vspace{2mm}

    \begin{minipage}{0.49\linewidth}
        \includegraphics[width=0.9\linewidth]{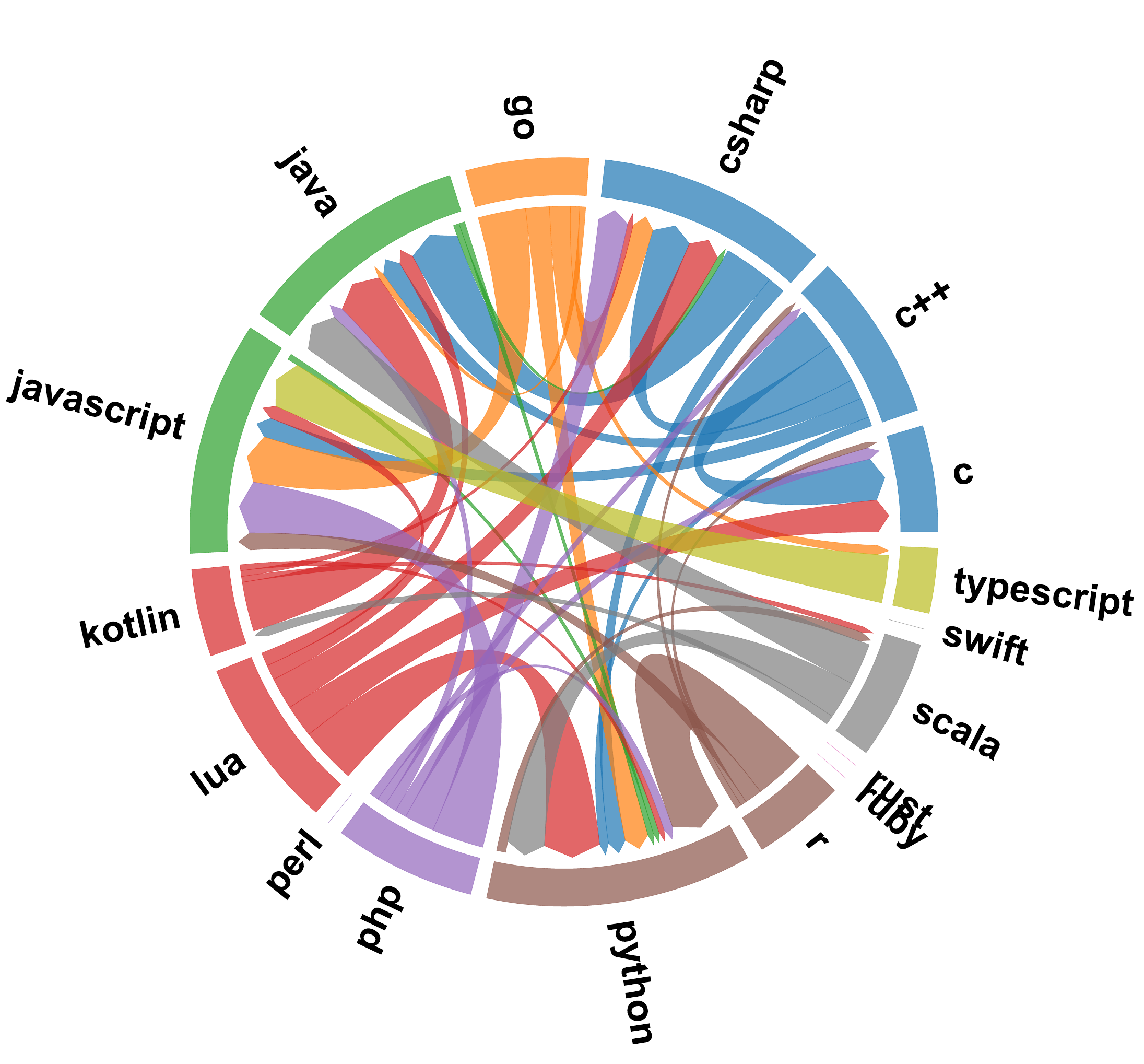}
        \centering
        \textbf{\footnotesize (c)~Llama-3.1-8B-Instruct}
    \end{minipage}
    \begin{minipage}{0.49\linewidth}
        \includegraphics[width=0.9\linewidth]{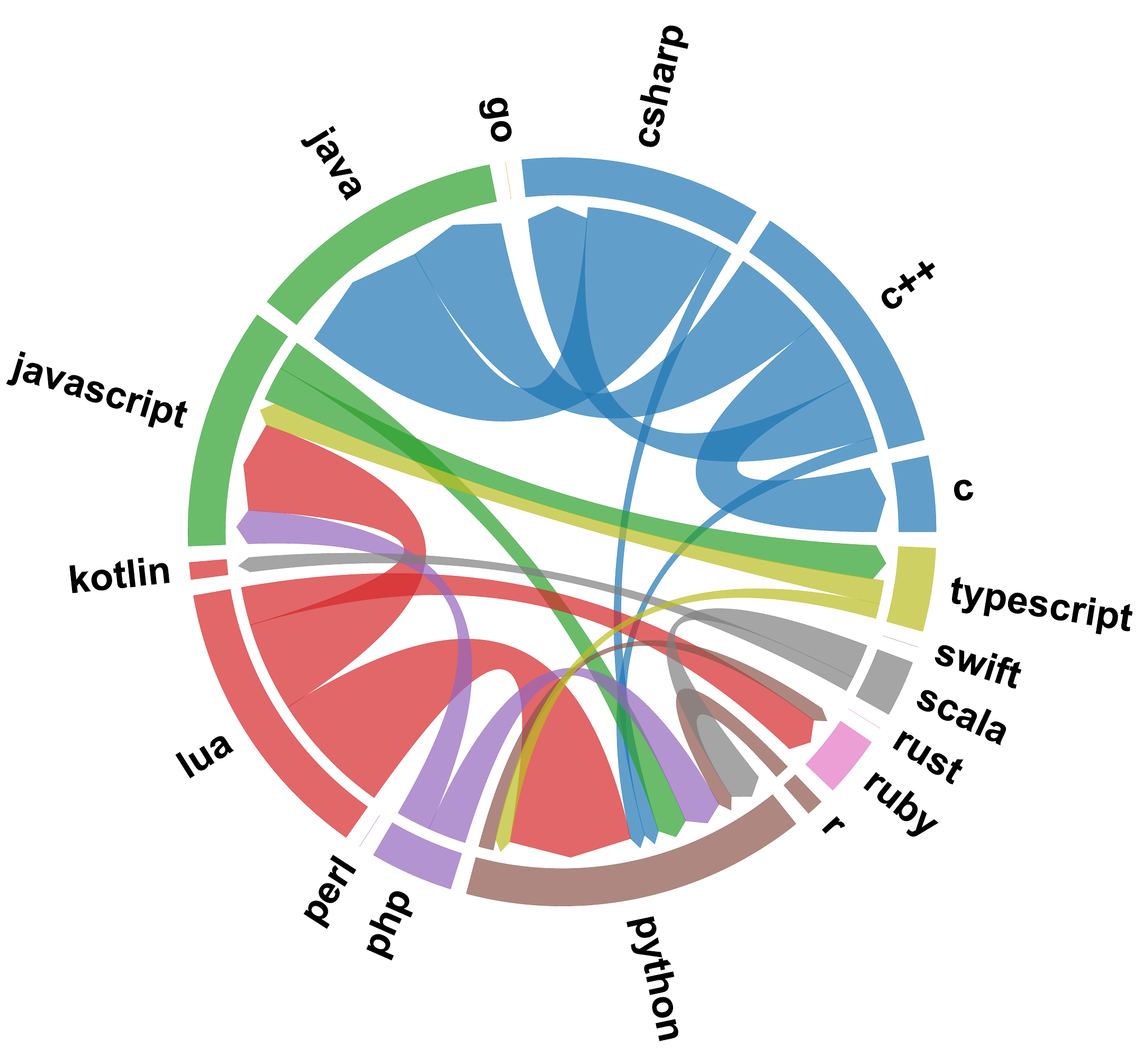}
          \centering
      \textbf{\footnotesize (d)~GPT-3.5-Turbo}
    \end{minipage}

    \caption{Migration Patterns in Programming Language Confusion during Code Generation (with BabelCode dataset). \normalfont \em \textbf{NOTE}: Data for all models are provided in the Supplementary file.}
    \label{fig:babelcode}
    \vspace{-0.4cm}
\end{figure}

We visualize the directional analysis of language migration in \Cref{fig:babelcode}. These chord diagrams reveal how PLC occurs, highlighting distinct migration patterns:
\begin{itemize}[leftmargin=*,noitemsep,topsep=2pt]
    \item Python as Universal Attractor: Python serves as the dominant destination language (high DMR) across confused models, reinforcing its status as the primary default schema.
    \item Family-Based Confusion: Consistent migration occurs between structurally similar languages, such as C\#/C++ and Java, and PHP/TypeScript shifting to JavaScript.
    \item Diffuse Migration: Languages like Lua, Kotlin, and Scala exhibit highly diffuse migration patterns, suggesting they are the least tethered to distinct internal representations and thus the most prone to incorrect generation.
\end{itemize}
Models with higher LCPR exhibit fewer and thinner migration ribbons, indicating more selective confusion, while low-LCPR models show widespread, distributed confusion, suggesting a failure to maintain clear internal language boundaries.

\subsubsection{Impact of Natural Language and Explicit Specification}
\label{results:humaneval}
The HumanEval-XL results (Table \ref{tab:h_lang_performance}), derived from prompts containing explicit language keywords, show a dramatic improvement in fidelity compared to the multilingual evaluation on BabelCode. This demonstrates that explicit language specification is highly effective at constraining model output and mitigating the general PLC observed previously. 
However, two secondary findings are established. When the target language is explicitly specified, nearly all models achieve $\text{LCPR} > 99.00\%$, with several reaching $100.00\%$. This confirms that their core competency to adhere to explicit instructions is robust.
Despite instructions spanning 23 different natural languages, the variation between the ``English" and ``Non-English" LCPR is negligible (generally $< 3\%$). This reinforces the conclusion that the model's tendency to confuse programming languages stems from deeper architectural or training data limitations, rather than difficulties interpreting the natural language of the prompt.

Overall, our analysis reveals six key findings:
\begin{itemize}[leftmargin=*]
    \item \textbf{Finding 1.1: PLC is pervasive without explicit cues, regardless of specialization.} Even code-specialized LLMs show significant confusion on the multi-lingual BabelCode task. Surprisingly, specialization does not guarantee superior language fidelity.
    \item \textbf{Finding 1.2: Syntactic correctness is preserved despite language confusion.} LLMs prioritize structural validity, leading to a systematic language migration towards syntactically familiar languages (e.g., Python, JavaScript) when uncertain.
    \item \textbf{Finding 1.3: High CPPR masks critical functional failure, yet some correctness remains.} While language confusion introduces semantic errors, leading to a substantial degradation in FPR, models still produce code that passes at least one unit test in confused samples. This suggests that the confusion does not always result in a total functional failure.
    \item \textbf{Finding 1.4: Strong directional biases exist in language confusion patterns.} Models predominantly migrate code to Python and between syntactically similar language families.
    \item \textbf{Finding 1.5: Explicit language keywords are the most effective mitigation for PLC.} Introducing explicit programming language keywords into the prompt significantly improves LCPR across all models.
    \item \textbf{Finding 1.6: Natural language instruction variations have minimal impact on PLC.} The instruction's natural language (e.g., Spanish vs. English) does not drive the confusion; the issue is the internal instability of language boundaries.
    \item \textbf{Finding 1.7: Code specialization does not guarantee language fidelity.} Specialized code models such as StarCoder2-15B-Instruct, DeepSeek-Coder-V2-Lite-Instruct-16B, and CodeLLama-13B-Instruct—despite superior code generation capabilities—exhibit higher PLC rates than smaller general-purpose models like Mistral-7B-Instruct. This suggests that code quality and language fidelity may require different optimization strategies.
\end{itemize}

\find{{\bf [RQ-1]} LLMs consistently confuse programming languages without explicit specification, exhibiting strong biases toward Python and between syntactically similar languages. While explicit language keywords drastically reduce confusion, the choice of natural instruction language has a negligible effect. Though syntactic correctness is preserved (systematic migration), this masks a severe degradation in functional correctness, even if some confused samples still pass minimal tests. Model size does not predict language fidelity, as specialized code models often show higher confusion rates than smaller general-purpose models.}

\begin{table*}[h]
\centering
\centering
    \caption{Language Confusion Pass Rates on HumanEval-xl prompts (Data for all the Natural Languages table is available in the Supplementary file)}
    \renewcommand{\arraystretch}{1.2}
\resizebox{\textwidth}{!}{%
\begin{tabular}{|l|c|c|c|c|c|c|c|c|c|c|}
\hline
\addlinespace[2pt]
Language 
& \makecell{CodeLLaMA \\ 13B-Instruct} 
& \makecell{DeepSeek-Coder-V2 \\ Lite-Instruct-16B} 
& \makecell{DeepSeek-V2-Lite \\ Instruct-16B} 
& \makecell{GPT-3.5 \\ Turbo} 
& \makecell{GPT-4.1 \\ Mini} 
& \makecell{LLaMA-3.1-8B \\ Instruct} 
& \makecell{Mistral-7B \\ Instruct} 
& \makecell{Qwen2.5 \\ Instruct-14B} 
& \makecell{Qwen2.5-Coder \\ Instruct-14B} 
& \makecell{Starcoder2 \\ 15B-Instruct} \\
\hline

English  & 96.10 & 99.79 & 99.36 & 100.00 & 99.89 & 99.89 & 99.79 & 100.00 & 100.00 & \cellcolor{lowcolor}95.53 \\
Non-English & 95.96 & 99.54 & 98.90 & 99.82 & 99.98 & 99.50 & 99.45 & 99.98 & 99.99 & \cellcolor{lowcolor}92.96 \\
All & 95.97 & 99.55 & 98.92 & 99.83 & 99.97 & 99.51 & 99.47 & 99.98 & 99.99 & \cellcolor{lowcolor}93.07 \\
\hline
\end{tabular}
}
\vspace{-0.4cm}
\label{tab:h_lang_performance}
\end{table*}

\subsection{[RQ2] Programming Language Confusion in LLM-Based Code Translation}
\label{rq2}

\noindent
\textbf{Goal.} We investigate how programming language confusion manifests during code translation tasks, examining whether models maintain target language adherence when explicitly tasked with converting code between programming languages.

\noindent
\textbf{Experiments.} We evaluated all models on the McEval dataset, measuring the language fidelity rate of the translated code. We analyzed performance variations across different source and target language pairs.

\noindent
\textbf{Results.} \Cref{tab:codenet} presents the overall results.
\noindent
\subsubsection{Translation Task Mitigates General PLC}
A primary observation is the significantly higher overall LCPR observed in the translation task compared to the multilingual code generation task. In \Cref{rq1}, average LCPR ranged from $74.33\%$ to $97.60\%$; in RQ2, LCPR is consistently high, ranging from $98.22\%$ to $99.73\%$ (\Cref{tab:codenet}).

This result indicates that the explicit requirement to handle two distinct programming languages (source and target) forces the LLMs into a different, more constrained operational mode, drastically reducing the propensity for the catastrophic, non-adherence errors seen in simple generation. The strong, explicit language tags in the translation prompt (e.g., ``Translate this code to Python") appear to serve as a mandatory constraint, mitigating the implicit language biases that emerged when the target language was the only instruction + function signature.

\noindent
\subsubsection{High Fidelity is Not Universal}
While overall LCPR is high, we observe performance plateaus and surprising rankings:
\begin{itemize}[leftmargin=*,noitemsep,topsep=2pt]
    \item \textbf{Generalist Models Excel:} GPT-3.5-Turbo and GPT-4.1-Mini achieve nearly perfect fidelity, suggesting that strong general instruction-following capabilities are paramount in this two-language, explicitly constrained setting.
    \item \textbf{Specialization Does Not Guarantee Control:} Models highly specialized in code, such as CodeLLaMA-13B-Instruct ($99.02\%$ LCPR) and Starcoder2-15B-Instruct ($98.22\%$ LCPR), lag behind models like Qwen2.5-Coder-Instruct ($99.73\%$ LCPR) and even the general-purpose GPT models. This reinforces the finding from RQ1 that code specialization does not necessarily translate to superior control over language boundaries or instruction adherence.
    \item \textbf{The Worst-Case Scenario is Rare:} The PLC observed in code translation is minimal (less than $2\%$ for most models). This suggests that confusion in translation is not a failure of identifying the target language, but likely a failure to escape the source language schema or a fallback to a dominant language when translation itself fails (i.e., the model outputs the source language or a dominant language like Python/C++ instead of the target).
\end{itemize}

\noindent
\subsubsection{ Minimal Impact of Natural Language}
The analysis of language fidelity based on the natural language of the surrounding instruction further confirms the finding from RQ1. The difference in LCPR between English and Non-English instructions is minimal across all models (generally $< 0.4\%$). This confirms that in highly explicit, structured coding tasks like translation, the model's fidelity is driven by the programming language tokens and syntax in the prompt.

\noindent
Overall, our analysis reveals four key findings:
\begin{itemize}[leftmargin=*]
    \item \textbf{Finding 2.1: Explicit Translation Mitigates General PLC.} The highly constrained nature of the code translation task drastically reduces the wide-ranging language confusion observed during general code generation. The requirement to process a source language acts as a strong anchor, constraining the model's output to the explicit target language.
    \item \textbf{Finding 2.2: Translation Fidelity is Robust (but not perfect).} LCPR is consistently high ($>98\%$) for the translation task. When confusion does occur, it is a rare failure mode, likely rooted in the model's inability to fully switch from the source language schema or a total failure leading to a default language fallback.
    \item \textbf{Finding 2.3: Specialization is not a Translation Advantage.} General-purpose models (GPTs variants) often achieve higher translation fidelity than larger, code-specialized models, confirming that adherence to complex, two-part instructions (source $\to$ target) is more about general robustness than code training volume alone.
    \item \textbf{Finding 2.4: Natural Language in Translation.} As with code generation, the natural language of the instruction has a negligible impact on translation fidelity. The model is constrained primarily by the explicit programming language keywords.
\end{itemize}

\find{{\bf [RQ-2]} Programming Language Confusion is minimal in the explicit code translation task, with LCPR consistently above $98\%$ across all models. The need to process a source language drastically mitigates the PLC observed in general generation. However, some specialized models still fail more often than generalist ones, confirming that explicit constraints and general instruction adherence are the dominant factors, not code specialization. }
\vspace{-0.2cm}

\begin{figure}[h]
\vspace{-0.4cm}
    \centering
        \begin{minipage}{0.49\linewidth}
        \includegraphics[width=\linewidth]{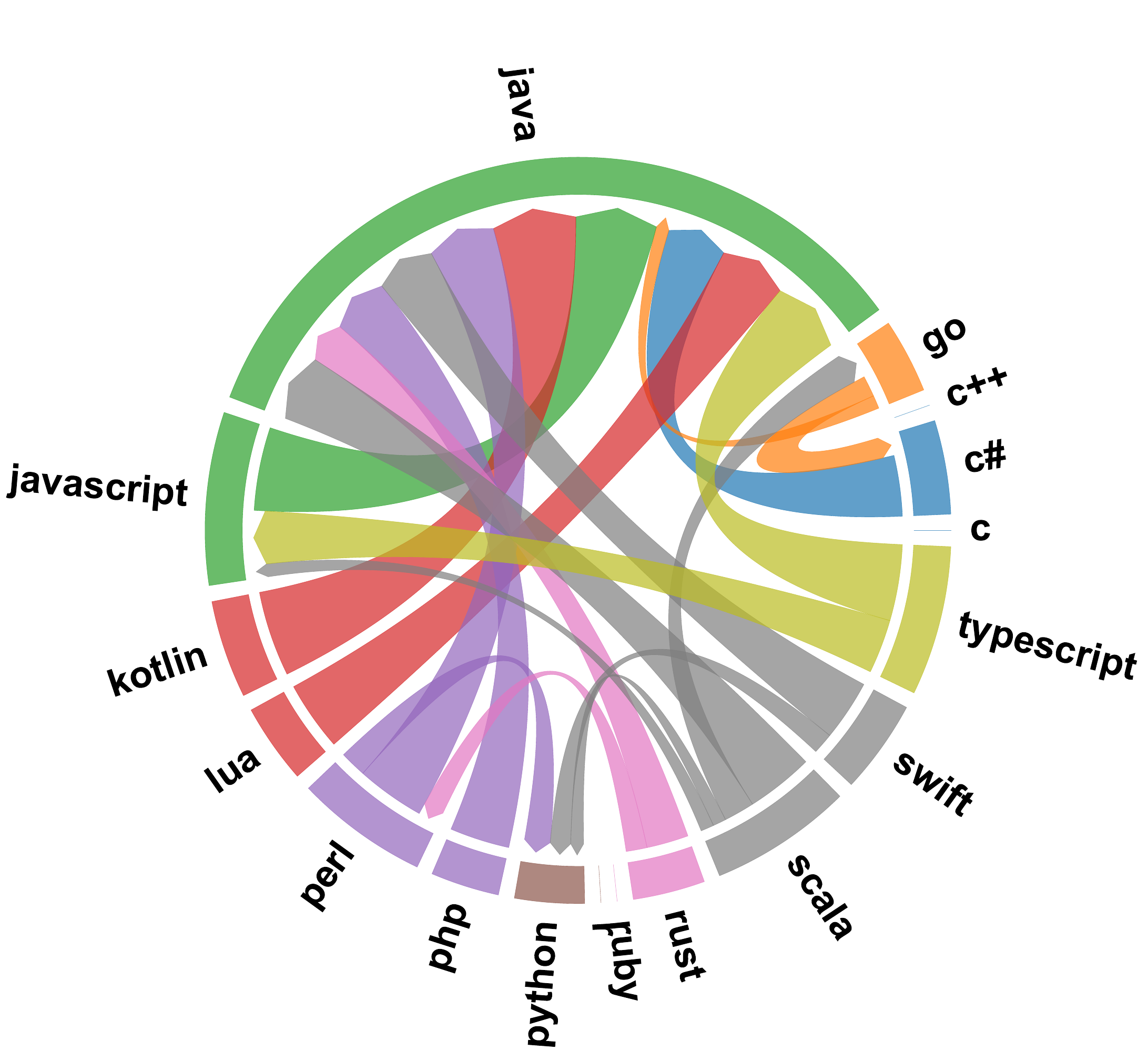}
        \centering
        \textbf{\small (a) Starcoder2-15B-Instruct (Source PL $\to$ Confused PL)}
    \end{minipage}
    \begin{minipage}{0.49\linewidth}
        \includegraphics[width=\linewidth ]{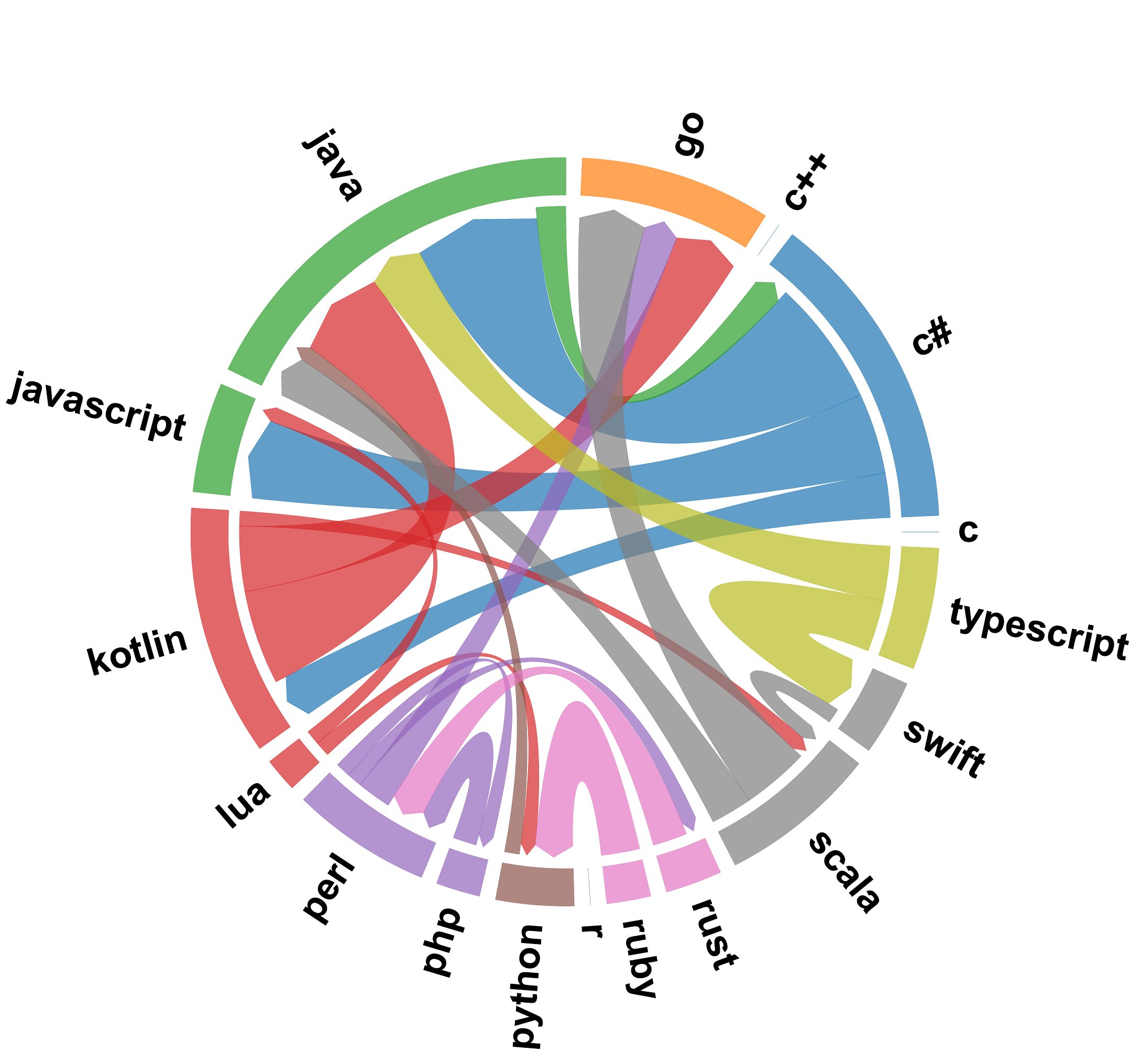}
        \centering
        \textbf{\small (b) Starcoder2-15B-Instruct (Target PL $\to$ Confused PL)}
    \end{minipage}
    
    \begin{minipage}{0.49\linewidth}
        \includegraphics[width=\linewidth ]{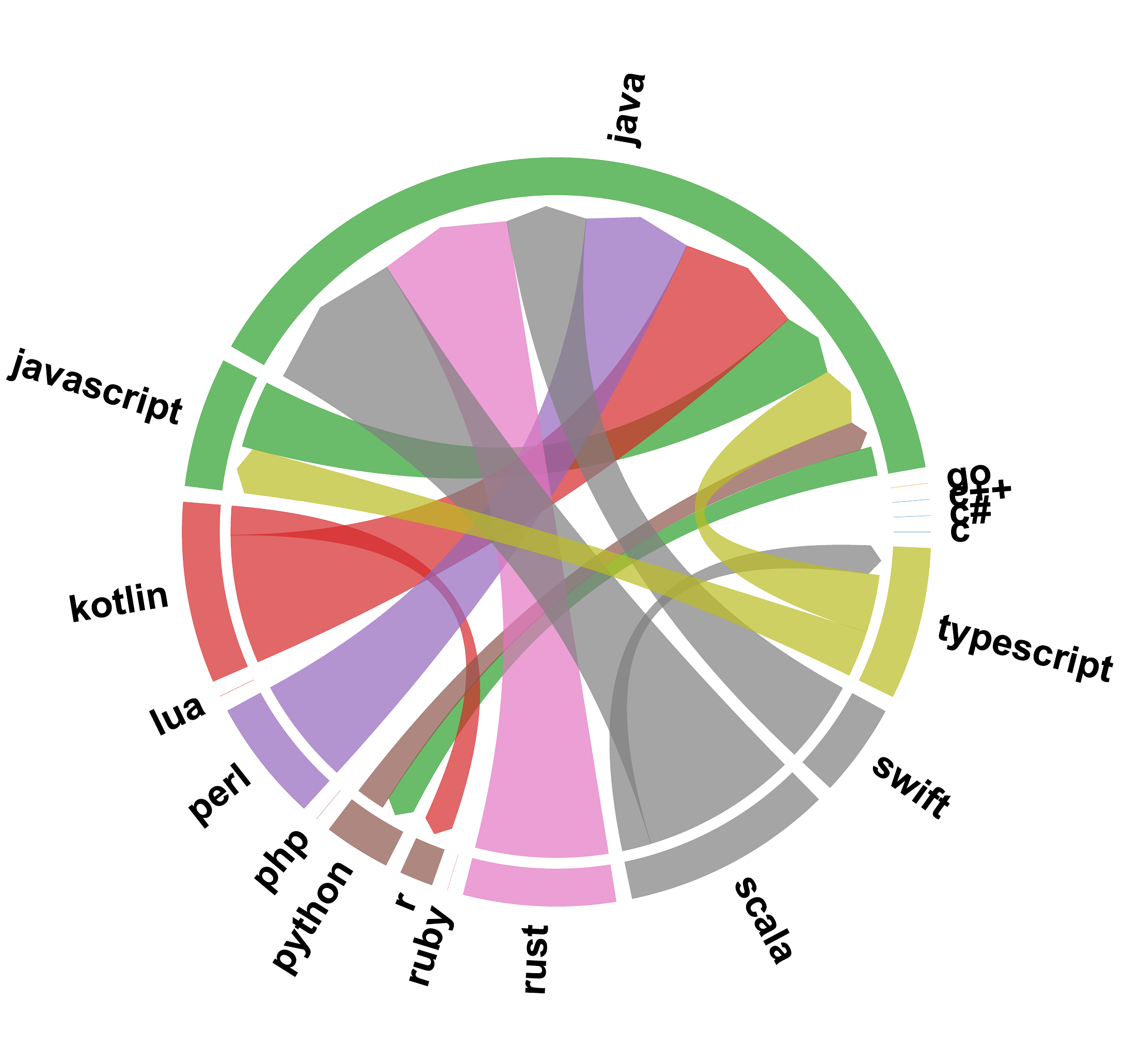}
        \centering
        \textbf{\small (c) CodeLLama-13B-Instruct \\(Source PL $\to$ Confused PL)}
    \end{minipage}
    \begin{minipage}{0.49\linewidth}
        \includegraphics[width=0.9\linewidth ]{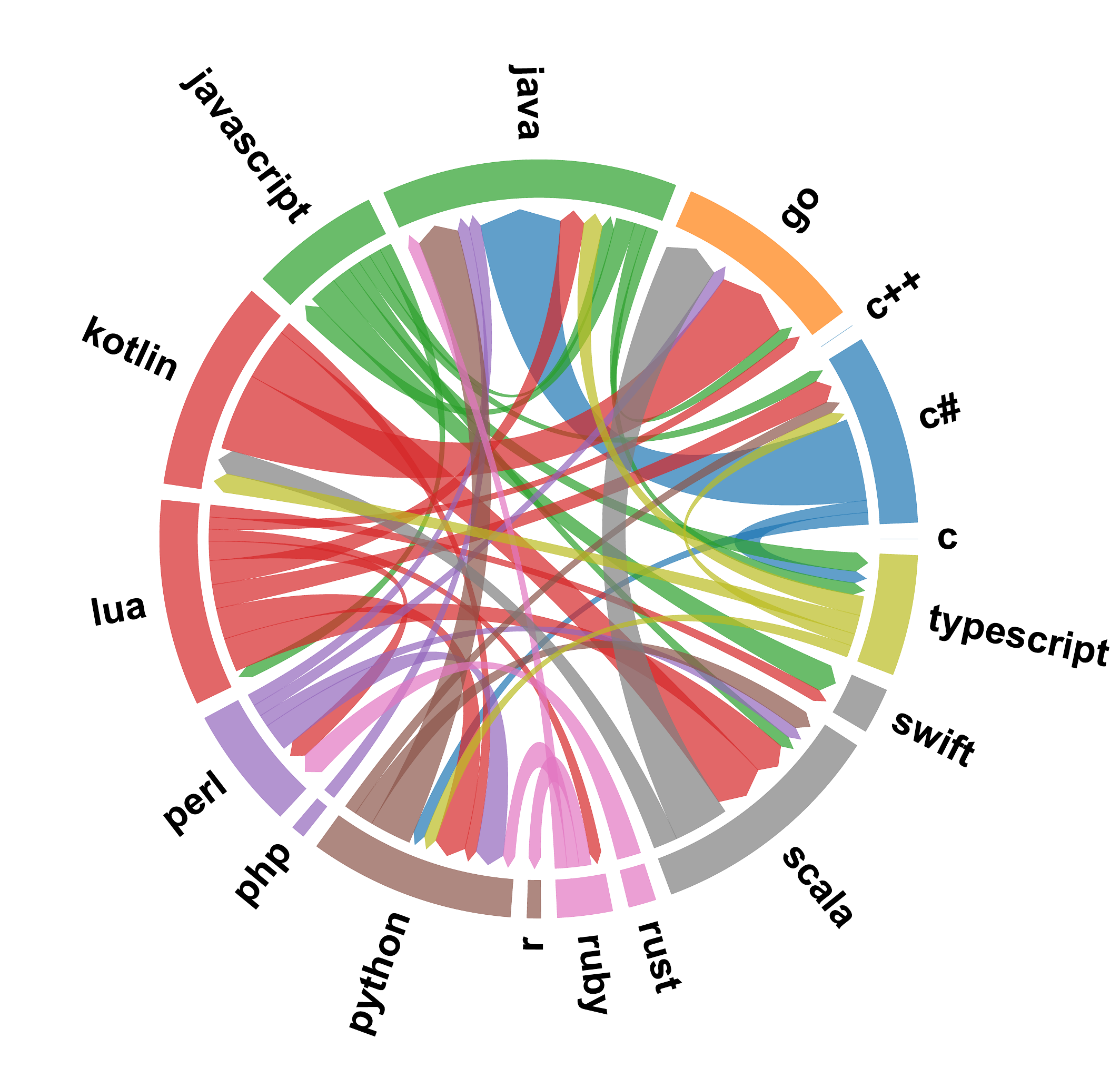}
        \centering
        \textbf{\small (d)~CodeLLama-13B-Instruct \\(Target PL $\to$ Confused PL ) }
    \end{minipage}
        
\caption{Directional Analysis of Programming Language Confusion in Translation Tasks. \\
    \normalfont \em \textbf{NOTE}: Data for all models are provided in the Supplementary file.}    \label{fig:translation_direction}
\vspace{-0.4cm}
\end{figure}

\begin{table*}[h]
\centering
\centering
    \caption{Language Confusion Pass Rates by Natural Languages during Code Translation with McEval}
    \renewcommand{\arraystretch}{1.2}
\resizebox{\textwidth}{!}{%
\begin{tabular}{|l|c|c|c|c|c|c|c|c|c|c|}
\hline
\addlinespace[2pt]
Language 
& \makecell{CodeLLaMA \\ 13B-Instruct} 
& \makecell{DeepSeek-Coder-V2 \\ Lite-Instruct-16B} 
& \makecell{DeepSeek-V2-Lite \\ Instruct-16B} 
& \makecell{GPT-3.5 \\ Turbo} 
& \makecell{GPT-4.1 \\ Mini} 
& \makecell{LLaMA-3.1-8B \\ Instruct} 
& \makecell{Mistral-7B \\ Instruct} 
& \makecell{Qwen2.5 \\ Instruct-14B} 
& \makecell{Qwen2.5-Coder \\ Instruct-14B} 
& \makecell{Starcoder2 \\ 15B-Instruct} \\
\hline
English  & 99.01 & 99.79 & 98.80 & 99.65 & 99.23 & 99.37 & 99.93 & 99.37 & 99.65 & \cellcolor{lowcolor}98.59 \\
Non-English  & 99.02 & 99.60 & 98.76 & 99.61 & 99.31 & 99.22 & 99.43 & 99.44 & 99.74 & \cellcolor{lowcolor}98.20 \\
All &  99.02 & 99.61 & 98.76 & 99.61 & 99.30 & 99.22 & 99.45 & 99.43 & 99.73 & \cellcolor{lowcolor}98.22 \\
\hline
\end{tabular}
}
\label{tab:codenet}
\vspace{-0.4cm}
\end{table*}

\subsection{[RQ3] How does model quantization affect programming language confusion in LLM-based code tasks?}
\label{rq3}

\noindent
\textbf{Goal.} We investigate how model quantization impacts the ability of LLMs to maintain programming language fidelity during code translation. We quantify PLC in quantized models and compare it to their full-precision counterparts to understand the effects of reduced numerical precision on language preservation and code correctness.

\noindent
\textbf{Motivation for Quantization Comparison.} Quantized models (e.g., Q4\_K\_M or Q5\_K\_M via frameworks like Ollama \footnote{https://ollama.com/}) are crucial for practical, local deployment and edge computing of LLMs. We aim to determine the real-world fidelity penalty incurred by reducing model precision for accessibility. Comparing these Quantized (Q) models against their Full-Precision (FP) counterparts evaluated in RQ1 and RQ2 allows us to isolate the degradation specific to the quantization process.

\noindent
\textbf{Experiments.} We conduct code generation and translation experiments using all datasets from RQ1 and RQ2 for a subset of models, focusing on CodeLLaMA-13B-Instruct, LLaMA3.1-8B-Instruct, Mistral-7B-Instruct, and StarCoder2-15B-Instruct, obtained from the Ollama framework. Quantized versions (\Cref{tab:ollama}) were evaluated alongside their full-precision counterparts (\Cref{tab:babelcode,tab:codenet}) to assess the difference in PLC, syntactic correctness, and language migration patterns.

 % Placeholder for the table content
\noindent
\textbf{Results.} Our analysis reveals several key patterns.
\noindent
\subsubsection{Quantization Significantly Increases PLC}
Comparing the LCPR values for the BabelCode dataset reveals a substantial drop in language fidelity for the quantized models, confirming the findings from the previous experiments. CodeLLaMA-13B-Instruc's LCPR drops from $\sim 87.0\%$ (FP) to $73.57\%$ (Q).
Similarly, StarCoder2-15B-Instruct's LCPR drops dramatically from $\sim 74.3\%$ (FP) to $51.87\%$ (Q).
This reduction in fidelity is also evident in the HumanEval-XL dataset. For StarCoder2-15B-Instruct, the LCPR drops from $\sim 93.07\%$ to $68.93\%$ (Q). This indicates that the information crucial for maintaining explicit constraints is also sensitive to numerical precision loss. Quantization effectively makes models ``forget" the finer distinctions between languages, amplifying the intrinsic confusion tendencies found in RQ1, even when explicit guidance is given.

\subsubsection{Quantization Impacts Default Bias and Migration Patterns}
\begin{figure}[h]
    \centering
        \begin{minipage}{0.49\linewidth}
        \includegraphics[width=\linewidth]{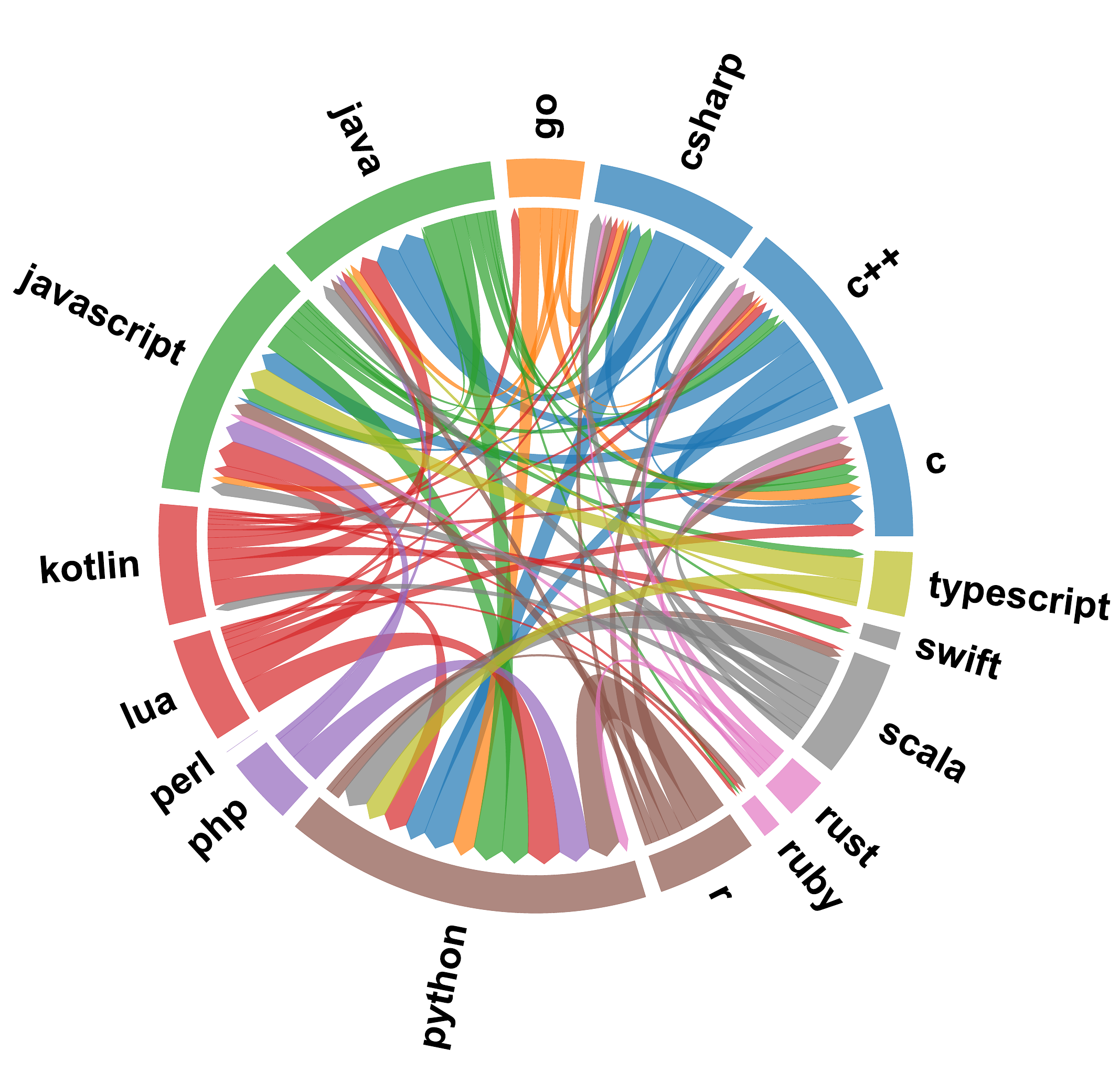}
        \centering
        \textbf{\small (a) CodeLLama-13B-Instruct-Q \\(BabelCode)}
    \end{minipage}
    \begin{minipage}{0.49\linewidth}
        \includegraphics[width=\linewidth ]{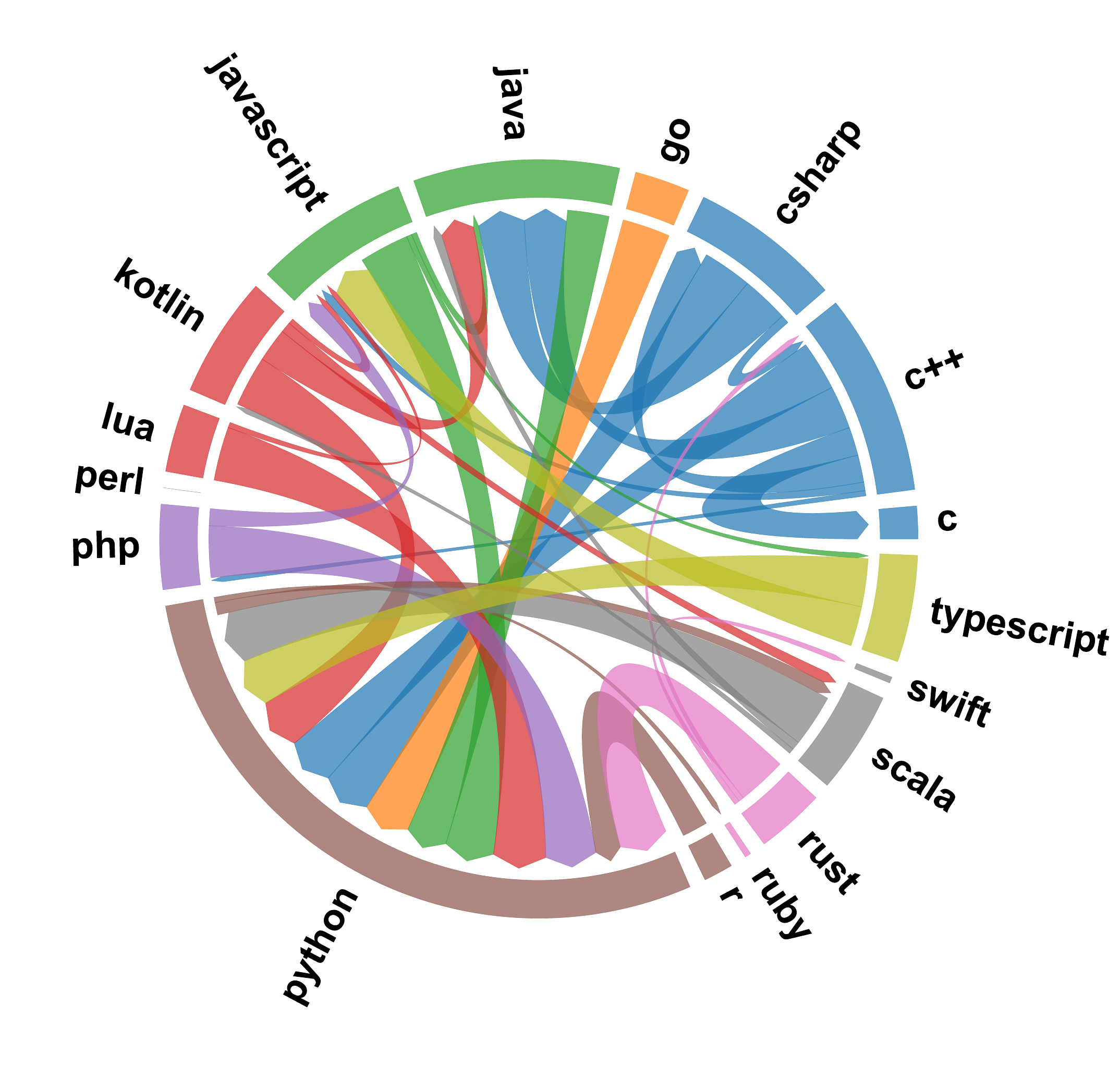}
        \centering
        \textbf{\small (b) Starcoder2-15B-Instruct-Q \\(BabelCode)}
    \end{minipage}
    
    \begin{minipage}{0.49\linewidth}
        \includegraphics[width=\linewidth ]{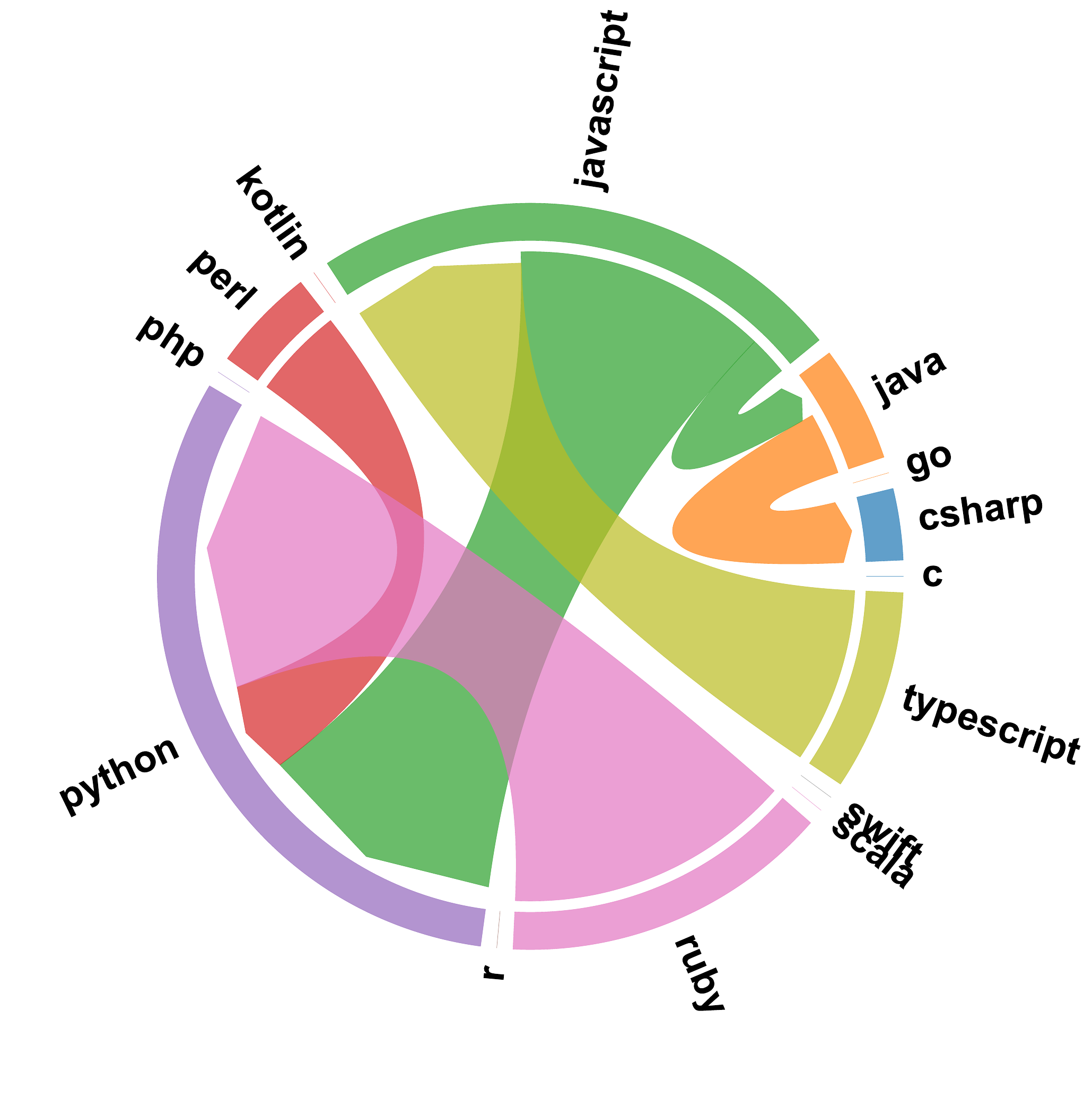}
        \centering
        \textbf{\small (c) CodeLLama-13B-Instruct-Q \\(HumanEval-XL)}
    \end{minipage}
    \begin{minipage}{0.49\linewidth}
        \includegraphics[width=0.9\linewidth ]{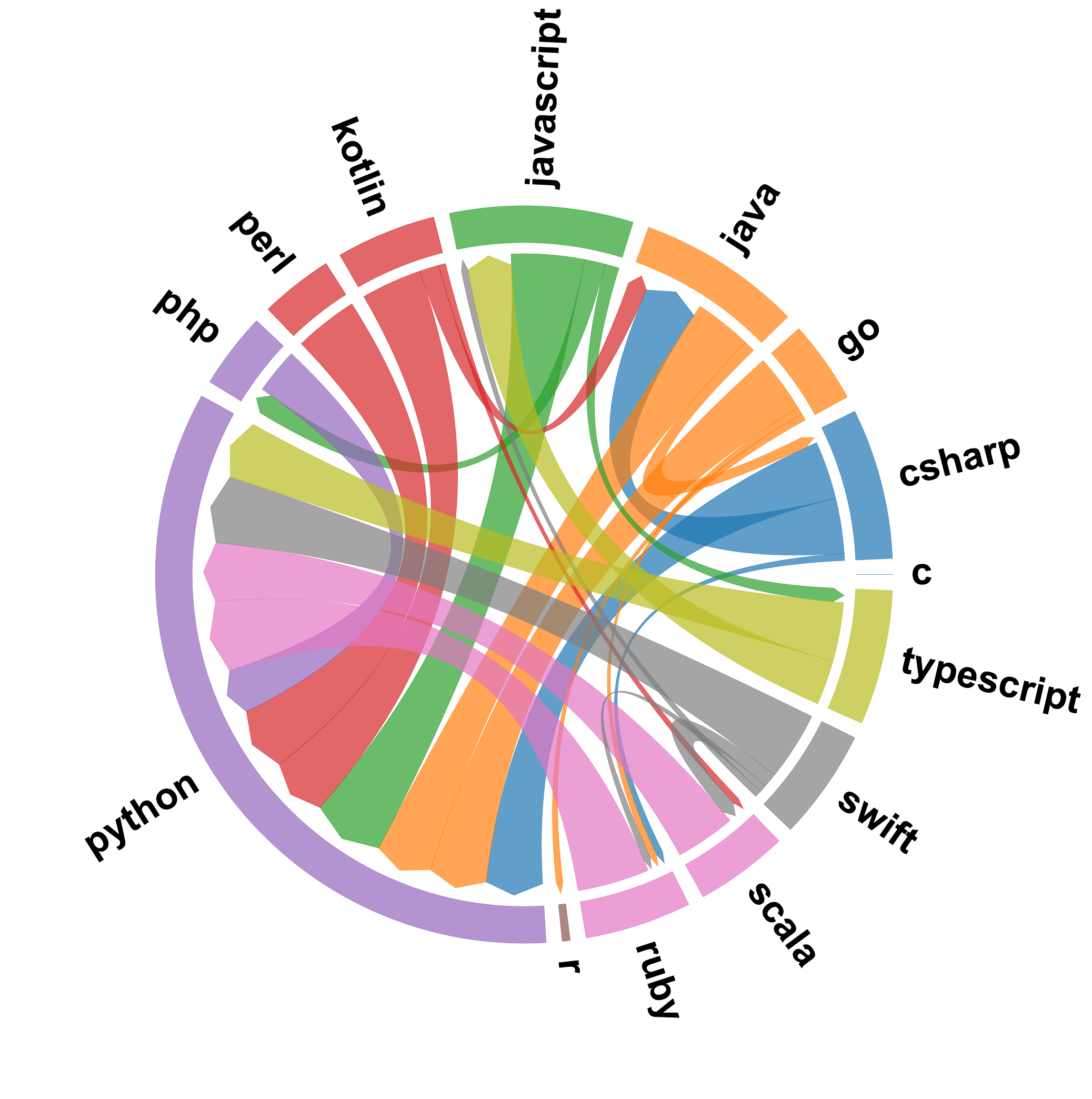}
        \centering
        \textbf{\small (d)~Starcoder2-15B-Instruct-Q \\(HumanEval-XL) }
    \end{minipage}
        
\caption{Directional Analysis of Programming Language Confusion when using quantized models}    \label{fig:translation_direction_ollama}
\vspace{-0.5cm}
\end{figure}

The analysis of the \Cref{fig:translation_direction_ollama} and DMR values reveals how quantization affects the models' internal language schemas:
\begin{itemize}[leftmargin=*]
    \item {Shift in Default Language:} For the generation task with no explicit language (LiveCodeBench), the quantization process can shift the model's default bias. The FP models generally defaulted to Python (RQ1), but the Q version of LLaMA3.1-8B-Instruct now exhibits a strong bias towards $C++$ ($89.55\%$ DMR). This shows that quantization does not just weaken boundaries; it can rearrange the internal language hierarchy.
    \item {Diffuse Migration:} Consistent with the qualitative analysis of the chord diagrams (See Appendix), the DMR values for BabelCode are slightly lower for the Q models (e.g., Mistral-7B-Instruct drops from $\sim 54\%$ to $41.43\%$). This suggests a qualitative change in confusion: instead of simply falling back to the strongest learned bias (Python), the model's output is scattered to a {more diverse and diffuse array of incorrect languages}, as precision loss weakens the target representations.
\end{itemize}

Overall, our analysis reveals four key findings:
\begin{itemize}[leftmargin=*]

    \item \textbf{Finding 3.1: Quantization is a significant amplifier of PLC.} Reducing model precision drastically increases PLC across all tasks, with LCPR dropping by up to $20$ percentage points, even when explicit language constraints are given.
    \item \textbf{Finding 3.2: Precision Loss Weakens and Re-arranges Language Boundaries.} Quantization can cause the model to shift its default language bias and lead to a more diverse and diffuse migration pattern.
    \item \textbf{Finding 3.3: Quantization Reduces Functional Reliability.} Quantization lowers functional pass rates across both confused and non-confused samples (Tables 2 and 3 in the Appendix).
    \item \textbf{Finding 3.4: Practical LLM Deployment Trades Off Fidelity.} The necessary process of quantization for local deployment introduces a significant, quantifiable penalty on language fidelity and reliability, confirming that the high performance reported on FP models may not translate directly to practical, resource-constrained use cases.
\end{itemize}

\find{{\bf [RQ-3]} Model quantization significantly amplifies PLC across all tasks, with LCPR dropping substantially even when explicit constraints are provided. The loss of precision also causes a breakdown of syntactic stability, leading to a simultaneous failure of language fidelity and code correctness in confused outputs during the complex translation task.}

\begin{table}[!http]
    \centering
    \caption{PLC in \% using quantized models from Ollama}
    \renewcommand{\arraystretch}{1.3}
    \footnotesize
          \resizebox{1\linewidth}{!}{
                    \begin{tabular}{l|c|c|c|c|l}
                    \hline
                        \textbf{LLMs} & \textbf{\makecell{Dataset}} &\textbf{\makecell{LCPR}} & \textbf{\makecell{CPPR non-confuse} }  & \textbf{\makecell{CPPR confuse}} & \textbf{\makecell{DMR \%}} \\ \hline
                         \multirow{4}{*} {\vspace{-0.1cm} CodeLLama-13B-Instruct-Q }
                         & LiveCodeBench & 97.49 & 99.88 & 81.82 & 40.91 (c))\\
                         & Babelcode & 73.57 & 98.20 & 96.37 & 71.03 (py)\\
                         & HumanEval-XL & 84.15 & 88.50 & 97.61 & 94.00 (py) \\
                         & McEval & 98.91 & 81.91 & 63.01 & 30.06 (java) \\ \hline

                         \multirow{4}{*} {\vspace{-0.1cm} LLama3.1-8B-instruct-Q }
                         & LiveCodeBench & 92.36 & 98.89 & 91.04 & 89.55 (c++)\\
                         & Babelcode & 65.08 & 98.89 & 98.03 & 75.51 (py)\\
                         & HumanEval-XL & 98.72 & 88.52 & 94.87 & 91.21 (py) \\
                         & McEval & 99.67 & 81.46 & 55.05 & 72.48 (java) \\ \hline

                         \multirow{4}{*} {\vspace{-0.1cm} Mistral-7B-Instruct-Q }
                         & LiveCodeBench & 99.89 & 97.72 & 100.00 & 100.00 (c++)\\
                         & Babelcode & 86.05 & 97.71 & 97.57 & 41.43 (js)\\
                         & HumanEval-XL & 99.75 & 88.35 & 88.89 & 38.89 (js) \\
                         & McEval & 99.68 & 80.14 & 35.24 & 40.95 (js) \\ \hline

                         \multirow{4}{*} {\vspace{-0.1cm} StarCoder2-15B-Instruct-Q }
                         & LiveCodeBench & 99.66 & 99.66 & 100.00 & 66.67 (java)))\\
                         & Babelcode & 51.87 & 99.31 & 98.87 & 86.95 (py))\\
                         & HumanEval-XL & 68.93 & 92.40 & 99.36 & 94.77 (py) \\
                         & McEval & 98.14 & 85.51 & 50.00 & 66.45 (java) \\ \hline
                    \end{tabular}}
         \label{tab:ollama}
\end{table}

\section{Discussion}
\label{sec:discussion}
This section examines the implications of our findings and discusses limitations of our study.
\subsection{Key Insights}
Our study shows that language confusion in LLMs is complex and depends heavily on how the model is built, the task it's doing, and how it's deployed.

\noindent
\textbf{Smart Error: Prioritizing Syntax over Language.}
When an LLM gets confused about which language to use, the error isn't random. It's a systematic move where the model tries to keep the code syntactically correct (meaning the code looks valid) even if it's in the wrong language. This is supported by high rates of correct-looking, but wrong-language, code. When this confusion happens, models most often default to Python, which acts like a ``Universal Attractor." This fallback happens because Python is so common in the model's training data. This shows the model is stuck between two goals: following the language instruction and producing code that simply ``looks right."

\noindent
\textbf{The Hidden Cost of Using Smaller Models (Quantization Penalty).}
To run LLMs locally or on small devices, they are often quantized (simplified by reducing numerical precision). This process significantly increases language confusion. Reducing the model's precision weakens its memory of distinct programming languages, leading to two major problems: (1) a noticeable drop in how well the model sticks to the target language, and (2) in complex tasks like code translation, a complete failure of the output code's structure and syntax when it gets confused. This means the impressive scores of large, full-size models might not hold up when they are shrunk for real-world use.

\noindent
\textbf{The Prompt's Natural Language Doesn't Matter.}
Our analysis across multiple spoken languages shows that the language of the instruction (e.g., English, Spanish) has very little effect on whether the model gets confused about the programming language. The confusion comes from the model's internal technical knowledge of programming languages, not its ability to understand the prompt's natural wording. Therefore, efforts to improve reliability should focus on the technical details of the prompt, not its grammar or wording.

\subsection{Mitigation Strategies}

Based on these findings, we recommend three main ways to reduce Programming Language Confusion:

\noindent
\textbf{Technical and Redundant Prompt Design.}
Since explicit programming language keywords are the best way to prevent confusion, and this prevention is easily lost during quantization, prompts must be extra detailed. Use redundant and highly technical cues like including language-specific boilerplate, function definitions, and using specific formatting tags (e.g., \texttt{```python}) to act as strong boundaries for the model.

\noindent
\textbf{Test Quantized Models for Confusion.}
When choosing models for local or small-device deployment, don't just look at the performance of the full-size version. Because model behavior changes so much after being quantized, the final, reduced-size version must be thoroughly tested for its robustness against language confusion. Since simpler models can sometimes be more reliable than specialized ones in certain constrained tasks, a careful, task-specific evaluation of the quantized model is essential before deployment.

\subsection{Threats to Validity}
Several limitations should be considered when interpreting our results, along with the steps we took to mitigate them.

\noindent
\textbf{Model Selection and Generalizability.} Our study evaluated a selection of LLMs with varying architectures and sizes. While this provides a good range of current models, the field is evolving very quickly, meaning our findings might not apply perfectly to every new model released.

\noindent
\textbf{Language Identification Accuracy.} Our method for identifying the programming language relies on a multi-tool approach, combining four different tools. Although this setup greatly improves reliability, it is not perfect. To reduce the chance of misclassification, we used a weighted consensus algorithm that trusts the most accurate tool while getting validation from the others.

\noindent
\textbf{Limited Language and Task Scope.} While our study covered 16 programming languages and 23 natural languages across different task types (generation and translation), we did not test every possible language or task. We managed this limitation by choosing languages that represent diverse programming styles (like imperative, functional, and object-oriented) and included both widely-used and less common languages to ensure broad coverage within our testing limits.

\noindent
\textbf{Intra-Statement Confusion Detection.} We found cases where models mixed the syntax of multiple languages within a single line or code block. Since current tools cannot formally evaluate this very fine-grained confusion, our analysis focused on identifying overall syntactic errors. These errors, often resulting from mixing language constructs, partially helped us understand where this fine-grained confusion was occurring.

\noindent
Our conclusions are based on observing the model's output (the generated code, whether it parses correctly, and if it passes tests). We did not analyze the model's internal workings, such as its internal thought processes (logits) or the specific details of its training data composition. Therefore, our insights are drawn from correlations in the output data.
\section{Conclusion}
\label{sec:conclusion}
This study introduces Programming Language Confusion (PLC) as a critical and systematic failure mode in LLMs. 
Rather than random error, PLC reflects consistent language migrations—most often defaults to Python—driven by training biases that favor syntactic plausibility over functional correctness. 
Explicit language keywords and structured tasks effectively mitigate confusion, while natural language instructions have little influence.

Another critical finding for real-world application is the effect of model quantization. Quantization, a necessary step for local deployment, acts as a severe amplifier of PLC. It causes a sharp drop in language fidelity across the board and, crucially, leads to a simultaneous collapse of language adherence and syntactic quality in complex tasks like translation. This demonstrates that the high performance metrics of full-precision models are not reliable indicators of performance in resource-constrained environments.

To address these issues, development workflows must shift focus: functional correctness, not syntactic pass rates, must be the primary validation goal, as the latter is often misleading. Practitioners must use explicit technical markers in prompts to provide robust language boundaries. Finally, quantization-aware testing is essential before deployment, recognizing that practical LLM use requires rigorous verification of language stability. Future research must prioritize redesigning LLM architectures to better encode and stabilize programming language boundaries for reliable multilingual code generation.

\section*{Data Availability}
\label{sec:data_availability}
To promote transparency and facilitate reproducibility, we make our artifacts available to the community at: 
\begin{center}
\url{https://github.com/TruX-DTF/PLC}
\end{center}

\bibliographystyle{IEEEtran}
\bibliography{IEEEtran}

\end{document}